\documentclass[prb, letterpaper,superscriptaddress, showpacs, twocolumn]{revtex4}
\usepackage{color}
\usepackage{hyperref}
\usepackage{amsmath}
\usepackage{amssymb}
\usepackage{epsfig}
\usepackage{graphicx}
\usepackage{amsmath}
\usepackage{bbm}

\begin{document}

\title{Dynamical mean-field theory for bosons}
\author{Peter Anders}
\affiliation{Theoretische Physik, ETH Zurich, 8093 Z{\"u}rich, Switzerland}
\author{Emanuel Gull}
\affiliation{Department of Physics, Columbia University, New York, NY 10027, USA}
\author{Lode Pollet}
\affiliation{Theoretische Physik, ETH Zurich, 8093 Z{\"u}rich, Switzerland}
\author{Matthias Troyer}
\affiliation{Theoretische Physik, ETH Zurich, 8093 Z{\"u}rich, Switzerland}
\author{Philipp Werner}
\affiliation{Theoretische Physik, ETH Zurich, 8093 Z{\"u}rich, Switzerland}
\date{\today}

\begin{abstract}
We discuss the recently developed bosonic dynamical mean-field (B-DMFT) framework, which maps a bosonic lattice model onto the selfconsistent solution of a bosonic impurity model with coupling to a reservoir of normal and condensed bosons. The effective impurity action is derived in several ways: (i) as an approximation to the kinetic energy functional of the lattice problem, (ii) using a cavity approach, and (iii) by using an effective medium approach based on adding a one-loop correction to the selfconsistently defined condensate. To solve the impurity problem, we use a continuous-time Monte Carlo algorithm based on a sampling of a perturbation expansion in the hybridization functions and the condensate wave function. As applications of the formalism we present finite temperature B-DMFT phase diagrams for the bosonic Hubbard model on a 3d cubic and 2d square lattice, the condensate order parameter as a function of chemical potential, critical exponents for the condensate, the approach to the weakly interacting Bose gas regime for weak repulsions, and the kinetic energy as a function of temperature. 
\end{abstract}


\hyphenation{}

\maketitle

\section{Introduction}

Particles with bosonic statistics can macroscopically occupy a single mode at low enough temperature, even in the absence of correlation. This phenomenon is known as Bose-Einstein condensation, and leads to a variety of phases in strongly correlated bosonic systems.
A typical example is $^4$He, which exhibits normal, superfluid, solid, and possibly supersolid phases. 
Another example are dilute ultracold atomic gases, which are well described by a Bogoliubov Hamiltonian. Strong interaction effects can be induced by adding lasers producing an optical lattice. The resulting system is an essentially clean realization of the Bose-Hubbard model, which describes the competition between hopping and on-site repulsion. Commensurability effects in the lattice lead to a phase transition from a superfluid to a Mott insulator at integer fillings and strong enough interaction.~\cite{Fisher89} Both cold gases and  $^4$He can be controlled experimentally with great accuracy and are virtually free of impurities and disorder. 
Cold atomic gases have the additional flexibility of tunable interaction strengths, and provide the freedom of changing the mass by choosing different alkali atoms.

For both $^4$He and ultracold bosonic gases in an optical lattice, powerful numerical methods exist for the strongly-interacting regime. Path integral simulations based on the worm algorithm can sample up to $10,000$ bosonic atoms at $T=1$K for supersolid $^4$He, enabling the study (see Ref.~\onlinecite{Prokofev_review} for a review) of individual defects such as vacancies,\cite{Boninsegni06b} dislocations,\cite{Boninsegni07} and grain boundaries.\cite{Pollet07_grainboundary} For cold gases, up to $1.5\times 10^6$ atoms were studied~\cite{Pollet10_comment} and compared directly to experiment with excellent agreement in time-of-flight images.\cite{Trotzky10} 
Such simulations involve a stochastic evaluation of all connected and disconnected diagrams occurring in a high-temperature series expansion on a finite lattice, which is an expansion in the hopping or kinetic energy (over temperature) around the atomic limit.\cite{Prokofev98, Boninsegni06a, Pollet07_LOWA}

The Monte Carlo simulation of fermionic lattice models suffers from the notorious sign problem, which prevents the study of large systems in the most interesting parameter regime. A computationally tractable approximate method to simulate these models is dynamical mean-field theory (DMFT).\cite{Metzner89,Georges92,Georges96, Maier05, Kotliar06}
In these calculations, the many-body selfenergy is approximated by all local skeleton diagrams involving local propagators only, which implies a selfconsistent determination of the selfenergy and the local propagators. Non-local contributions are neglected. This simplification is convenient because the approximated selfenergy can be evaluated efficiently from an appropriately defined impurity action.\cite{Georges92} By using sign-free (for single-site DMFT) efficient continuous-time Monte Carlo solvers,\cite{Gull10} one obtains the full Green's function as a solution to the effective impurity action in polynomial time.\cite{Rubtsov05, Werner06, Gull08}  The simplification of the diagrammatic structure\cite{Metzner89} allows to define DMFT for arbitrary dimensions and lattice structures. 
A major success of DMFT is the understanding of the Mott metal-insulator transition it has provided.\cite{Georges96,Kotliar06} DMFT has been extensively used to study model systems and -- in conjunction with band structure techniques -- to compute material properties for a wide range of compounds.\cite{Kotliar06,Held07}
Several extensions make the approximation systematic and controlled: Cluster methods\cite{Maier05} like the dynamical cluster approximation\cite{Hettler98} or the cellular DMFT\cite{Lichtenstein00,Kotliar01} reintroduce momentum dependence by considering multi-site impurity clusters. Methods like D$\Gamma$A\cite{Toschi08} or dual fermions\cite{Rubtsov06,Hafermann08} systematically consider non-local diagrams beyond DMFT. In principle, a diagrammatic Monte Carlo evaluation of all neglected contributions to the selfenergy allows to estimate its accuracy, as, e.g., recently done for the Anderson localization problem.\cite{Pollet10_inc}


The formulation of an analogous dynamical mean-field theory for bosonic lattice models has proven difficult. 
On the one hand, from the perturbative diagrammatic point of view, the idea of retaining an infinite subclass of bare diagrams may seem dubious since standard Feynman diagrammatic expansions in $U$ fail notoriously for the Bose-Hubbard model due to Dyson's collapse argument: In the complex plane, the convergence radius is zero as all bosons would collapse to a single point for negative interactions, with an infinite negative energy in the thermodynamic limit. Hence, it seems impossible to think of meaningful diagrammatic expansions in $U$ for bosons, and only a few techniques are known to deal with this problem such as Kleinert's variational perturbation theory,\cite{Kleinert01} the cutting off of large field contributions,\cite{Meurice02} or the use of a sequence with appropriately chosen counter terms rendering an infinite convergence radius in the absence of phase transitions.\cite{Pollet10_antidyson} In the Baym-Kadanoff approach to DMFT, a subset of diagrams consisting of all contributions from local dressed propagators are retained. These skeleton diagrams may have radically different convergence properties than the bare series but those properties are essentially unknown. In the case of absolutely convergent series the bare and skeleton series are equivalent and physically meaningful, but due to Dyson's collapse argument for the bare series there is no guarantee that the skeleton approach will converge.

On the other hand, keeping knowledge of the series provenance and by using the effective action for a {\it single} site ({\it ie.,} restoring the action on the basis of the series), may still be worthwhile. In the normal phase the usual DMFT formalism can be applied in the same way as it is usually done for fermions. By using an effective action on a single site (which can be solved with the method of choice), most of the aforementioned convergence issues can be sidestepped. It is in fact only the occurrence of a condensate (which happens in the single particle channel) in the broken symmetry phase that poses a challenge in the development of a B-DMFT formalism (cf. Ref~\onlinecite{Chitra01}). 
We will define our B-DMFT theory at the one-loop level beyond this selfconsistently defined mean-field (condensate). As shown in App.~\ref{medium_sec}, our effective action for the impurity problem is the most general action for an impurity with a broken $U(1)$ symmetry and a one-loop correction.~\cite{NegeleOrland}  The hybridizations are then determined selfconsistently with the underlying lattice problem. The approximations involved can still be understood in the language of diagrams,
but a full interpretation in terms of Baym-Kadanoff functionals remains subjective in light of the asymptotic series expansions. We therefore prefer to use alternative derivation schemes in this work that do not rely on an expansion in the bosonic repulsion $U$.

In the limit of infinite coordination number the decoupling approximation for the Bose-Hubbard model becomes exact (see the Appendix of Ref.~\onlinecite{Fisher89}). This decoupling approximation is recovered in our action at the mean-field level, since the one-loop correction vanishes. 
The original B-DMFT paper by Byczuk and Vollhardt\cite{Byczuk08} 
(as well as subsequent work presented by Hu and Tong\cite{Hu09}) 
is based on the assumption that 
 in this limit the kinetic and potential energy contributions in the broken symmetry phase can still be comparable to each other, in apparent contrast to Ref.~\onlinecite{Fisher89}. Their work postulates a scaling Ansatz with different scalings for condensed and non-condensed bosons, leading to an effective action that is different from the one we shall describe. 
The authors of Ref.~\onlinecite{Hubener09} performed an expansion in the inverse of the coordination number around the limit of infinite coordination number ({\it i.e.} the static mean-field result of Ref.~\onlinecite{Fisher89}), but treated the condensate in a perturbative way only valid on the Bethe lattice. After the publication of our results,\cite{Anders10} they corrected for this and also derived a fully selfconsistent (and general) version of the B-DMFT formalism.~\cite{Snoek10a}

The virtue of extending the DMFT framework to interacting Bose systems lies in the fact that certain model systems otherwise not amenable to bosonic simulations, e.g. models with frustrated interactions, or Bose-Fermi mixtures can now be studied numerically. The quality of the approximation $\Sigma(k,\omega) \rightarrow \Sigma^{\rm skel}(\omega)[G_{\rm loc}]$ is system dependent and needs to be established on a case-by-case basis. We will see that the B-DMFT approximation is very good for the single-component Bose-Hubbard model in 3d, and thus may serve as a good starting point for more complicated systems.

In both the fermionic and bosonic versions of dynamical mean-field theory, the computationally challenging part is the solution of the impurity problem. For fermionic systems exact diagonalization,\cite{Caffarel94} semi-analytical resummation of diagrams,\cite{Keiter71,Pruschke89} quantum Monte Carlo,\cite{Hirsch86} and numerical renormalization group\cite{Bulla98} methods are in wide use for single-orbital models.
In recent years, significant progress has been made with the development of diagrammatic Monte Carlo impurity solver techniques, based on an expansion of the partition function in powers of the interaction\cite{Rubtsov05,Gull08} or the impurity-bath hybridization,\cite{Werner06,Werner06Kondo} allowing access to much larger impurity clusters, lower temperatures, and more general interactions.\cite{Gull10}

In this paper we present a detailed derivation of the B-DMFT equations and show how the impurity-condensate coupling must be chosen to obtain a consistent theory.
We quantify the errors introduced by the dynamical mean-field approximation for a system with finite coordination number by comparing with lattice Monte Carlo methods, and we describe the impurity solver proposed in Ref.~\onlinecite{Anders10} in such  detail that the implementation of the method becomes straightforward.

The paper is organized as follows: Sec.~\ref{sec:modelx} introduces the Bose Hubbard model, and in Sec.~\ref{B-DMFT_sec} we derive the B-DMFT formalism, which is summarized in Sec.~\ref{sec:DMFT_procedure}. In section \ref{solver_sec} we describe the diagrammatic Monte Carlo impurity solver. Section \ref{limits_sec} discusses solvable limits 
while B-DMFT results for interacting bosons on a Bethe and 3d simple cubic lattice are presented in Sec.~\ref{results_sec}. 
Sec.~\ref{conclusion_sec} provides a summary and outlook.


\section{The Bose-Hubbard model}
\label{sec:modelx}

We consider a model of spinless bosons on a lattice, described by the Bose-Hubbard Hamiltonian in standard lattice notation,

\begin{equation}
H = - t \sum_{\langle i,j\rangle} b_i^{\dagger}b_j +\frac{U}{2}\sum_i n_i (n_i - 1) - \mu \sum_i n_i,
\label{hamiltonian}
\end{equation}
where $t$ denotes the hopping amplitude, $U$ the on-site interaction and $\mu$ the chemical potential. Unless otherwise written, our unit will be the hopping amplitude, $t=1$.

This model has three phases: (i) a normal phase at high temperature, (ii) a Mott insulating phase at zero temperature and commensurate filling for $U \gg zt$, and (iii) a superfluid phase occurring at low temperature.
In the following, we will develop an effective theory which goes beyond static mean-field theory by including temporal (in imaginary time) fluctuations, but which is restricted to a single site and therefore contains no momentum fluctuations.

The local terms of the action on a single site are
\begin{equation}
S_{\rm int} = \int_0^{\beta} d\tau b_{\rm int} ^*(\tau) ( \partial_{\tau} - \mu) b_{\rm int} (\tau) + \frac{U}{2}   n_{\rm int} (n_{\rm int}-1).
\label{eq:action_internal}
\end{equation}
The subscript `int' denotes the internal degrees of freedom.
This local action correctly describes the physics of the system in case of zero hopping $(t = 0)$, when a factorization over each lattice site is exact. Hence it already contains the low-energy physics occurring deep in the Mott phase. At very high temperature the action is also accurate because, in terms of Feynman's path integrals, the world-lines describing the evolution of particles in imaginary time remain almost straight lines, i.e. few hopping processes are present. Our task is then to replace this action with an effective action that also describes the physics deep in the superfluid phase (when the interaction is weak, compared to the hopping amplitude), {\it i.e.}, in the regime where the Bogoliubov theory of the weakly interacting Bose gas applies.
In terms of the treatment of the weakly interacting Bose gas (WIBG) of Ref.~\onlinecite{CapogrossoSansone10}, there is reason to believe that such an accurate DMFT-like effective action exists: To leading and subleading order (in the interaction), the normal and anomalous selfenergies are momentum independent at zero frequency, which is precisely the DMFT approximation. 

Note that in Ref.~\onlinecite{CapogrossoSansone10} an explicit small symmetry breaking field was added, which introduced a gap in the spectrum (and removed IR divergences in leading orders). Although this violates the requirement that the spectrum of a superfluid should be gapless,~\cite{Nepomnyashchii78, Nepomnyashchii83} it was argued in Ref.~\onlinecite{CapogrossoSansone10} that the leading orders of thermodynamic observables are found on short-range distances and provided by Beliaev's diagrammatic technique, while for long-range physics (ie, the long-range wavelength fluctuations of the order parameter) one has to resort to Popov's hydrodynamic theory. Similar considerations hold for our effective action where the gap is not fixed but depends on the condensate $\phi$. The short-range physics can be computed, but the long-range fluctuations are not part of the theory. In particular, the explicit symmetry breaking still leads to a finite condensate density in two dimensions at finite temperature (see below, sec.~\ref{2d_sec}), which greatly simplifies the theory. However,  while the explicit but fixed symmetry breaking field of Ref.~\onlinecite{CapogrossoSansone10} made the superfluid-normal transition first order, we expect in our case a second order-transition (see below, sec.~\ref{sec:critical_exponents}) with non-universal critical exponents different from static mean-field theory because of the dynamical corrections in the hybridization terms which form a one-loop correction to mean-field theory.~\cite{NegeleOrland} 


\section{Derivation of the B-DMFT equations}
\label{B-DMFT_sec}

In this section we present the derivation of the B-DMFT equations (the action and the selfconsistency relations). 
For completeness we present alternative (but equivalent) derivations of the DMFT formalism such as a quantum cavity reasoning (Appendix~\ref{sec:cavity}) and an effective medium approach (Appendix~\ref{medium_sec}) in the Appendix.
Here we implement an expansion around the atomic limit, following almost literally  the lecture notes by A. Georges,~\cite{Antoine} and consider B-DMFT as an approximation to the kinetic energy functional. This derivation can to a large extent also be found in the supplementary material accompanying our previous Letter, Ref.~\onlinecite{Anders10}. The atomic reference system is interpreted as the impurity problem. We use a the coupling constant integration method and introduce source fields (Lagrange multipliers) to constrain the condensate field and the connected Green's function for the normal bosons to their physical values. By doing so, we avoid the collapse arguments associated with an expansion in $U$ mentioned in the introduction of this paper.

\subsection{Expansion parameter}
We introduce a parameter $\alpha \in [0,1]$ such that 
%
\begin{equation}
H_{\alpha} = \frac{U}{2} \sum_i n_i (n_i - 1) - \alpha t \sum_{\langle i,j \rangle} b_i^{\dagger} b_j.
\label{eq:antoine_alpha}
\end{equation}
When $\alpha$ = 0, the atomic limit is recovered and the partition function factorizes over all sites. When $\alpha=1$ the full hopping is recovered, and this is the model we are ultimately interested in.
\subsection{Source fields and constraining fields}

Constraining the normal/anomalous Green's functions and the condensate to specified values can be done by introducing conjugate source fields (Lagrange multipliers) in the action. In order to constrain the condensate to $\mathbf{\Phi}$ we introduce the source field $\mathbf{J}$, and analogous for the {\it connected} Green's function $\mathbf{G}_c$ with source field $\mathbf{\Delta}$. Throughout this document we use the Nambu notation in which $\mathbf{\Phi}^\dagger=(\phi^*, \phi)$, $\mathbf{J}^\dagger=(J^*, J)$ and the individual components of $\mathbf{G}_c$ and $\mathbf{\Delta}$ are given by
\begin{eqnarray}
\mathbf{G}_c(\tau) =  \left( \begin{array}{cc}
G_c(\tau)  & \tilde{G}_c(\tau) \\
\tilde{G}_c^{*}(\tau) &  G_c(-\tau)
\end{array} \right),
\end{eqnarray}
and
\begin{eqnarray}
\mathbf{\Delta}(\tau)= \left( \begin{array}{cc}
F(-\tau)  & 2K(\tau) \\
2K^{*}(\tau) & F(\tau)
\end{array} \right).
\end{eqnarray}
We can then explicitly write down the grand potential per site (there are $N_s$ sites) which is a functional of the source fields and also depends on the constraining fields, 
\begin{widetext}
\begin{eqnarray}
\Omega_{\alpha}[\mathbf{J}, \mathbf{\Phi}, \mathbf{\Delta}, \mathbf{G_c}] & = & -\frac{1}{N_s \beta} \ln \int \mathcal{D}[b^*, b] \exp \Bigg\{ \int_0^{\beta} d\tau \left( \sum_i b_i^* ( -\partial_{\tau} + \mu ) b_i - H_{\rm \alpha}[b^*, b] \right) \nonumber \\
{} & {} & + \int_0^{\beta} d\tau \sum_i \left( J^*(\tau) [ b_i(\tau) - \phi_i(\tau) ] + J(\tau) [b_i^*(\tau) - \phi_i^*(\tau) ] \right)\nonumber \\
{} & {} & + \int_0^{\beta} d\tau \int_0^{\beta} d\tau' \sum_i F(\tau - \tau') [ \delta b_i(\tau) \delta  b_i^*(\tau') + G_c(\tau - \tau') ] \nonumber \\
{} & {} & + \int_0^{\beta} d\tau \int_0^{\beta} d\tau' \sum_i K(\tau - \tau') [ \delta b_i^*(\tau) \delta  b_i^*(\tau') + \tilde{G}_c^*(\tau - \tau') ] \nonumber \\
{} & {} & + \int_0^{\beta} d\tau \int_0^{\beta} d\tau' \sum_i K^*(\tau - \tau') [ \delta b_i(\tau) \delta  b_i(\tau') + \tilde{G}_c(\tau - \tau') ] \Bigg\}.
\end{eqnarray}
\end{widetext}
Here $\delta b$ is the non-condensed part of the operator $b$ given by $b = \langle b \rangle + \delta b$.

\subsection{Atomic limit : impurity model}
Let us consider the atomic limit $\alpha = 0$. There the problem is local on every site with grand potential
\begin{widetext}
\begin{eqnarray}
\Omega_{0}[\mathbf{J_0}, \mathbf{\Phi}, \mathbf{\Delta_0}, \mathbf{G_c}] & = & -\frac{1}{N_s \beta} \ln \int \mathcal{D}[b^*, b] \exp \Bigg\{ \int_0^{\beta} d\tau \left( \sum_i b_i^* ( -\partial_{\tau} + \mu ) b_i - \frac{U}{2} n_i(n_i-1) \right) \nonumber \\
{} & {} & + \int_0^{\beta} d\tau \sum_i \left( J_0^*(\tau) [ b_i(\tau) - \phi_i(\tau) ] + J_0(\tau) [b_i^*(\tau) - \phi_i^*(\tau) ] \right)\nonumber \\
{} & {} & + \int_0^{\beta} d\tau \int_0^{\beta} d\tau' \sum_i F_0(\tau - \tau') [ \delta b_i(\tau) \delta  b_i^*(\tau') + G_c(\tau - \tau') ] \nonumber \\
{} & {} & + \int_0^{\beta} d\tau \int_0^{\beta} d\tau' \sum_i K_0(\tau - \tau') [ \delta b_i^*(\tau) \delta  b_i^*(\tau') + \tilde{G}_c^*(\tau - \tau') ] \nonumber \\
{} & {} & + \int_0^{\beta} d\tau \int_0^{\beta} d\tau' \sum_i K_0^*(\tau - \tau') [ \delta b_i(\tau) \delta  b_i(\tau') + \tilde{G}_c(\tau - \tau') ] \Bigg\}.
\end{eqnarray}
\end{widetext}
From $\delta\Omega_0/\delta J_0=0$ and $\delta\Omega_0/\delta J_0^*=0$ we obtain 
\begin{equation}
\mathbf{\Phi}=\langle\mathbf{b}\rangle_{S_{\text{imp}}},
\label{phi}
\end{equation}
and from $\delta\Omega_0/\delta F=0$, $\delta\Omega_0/\delta K=0$, $\delta\Omega_0/\delta K^*=0$ the relation 
\begin{equation}
\mathbf{G}_c(\tau)=\mathbf{G}(\tau)+\mathbf{\Phi}\mathbf{\Phi}^\dagger,
\label{Gc}
\end{equation}
with $\mathbf{G}(\tau)=-\langle T\mathbf{b}(\tau)\mathbf{b}^{\dagger}(0)\rangle_{S_\text{imp}}$. The expectation values $\langle \ldots \rangle_{S_\text{imp}}=Tr[T e^{-S_\text{imp}} \ldots ]/Tr[T e^{-S_\text{imp}}]$ are defined with respect to an impurity action
\begin{eqnarray}
S_{\rm imp} & = & - \frac{1}{2}\int_0^{\beta} \int_0^{\beta} d\tau  d\tau' \delta \mathbf{b}^*(\tau) {\mathbf{\Delta}}_0 (\tau - \tau')  \delta\mathbf{b}(\tau')\nonumber \\ 
& - & \mu \int_0^\beta d\tau n(\tau) + \frac{U}{2} \int_0^{\beta} d\tau n(\tau) [ n(\tau) - 1]\nonumber \\
&-& \int_0^{\beta} d\tau \mathbf{J}_0^{\dagger}(\tau) \mathbf{b}(\tau).
\label{imp_a}
\end{eqnarray}
Inverting expressions (\ref{phi}) and (\ref{Gc}) yields $\mathbf{J}[\mathbf{\Phi}, \mathbf{G_c} ]$ and $ \mathbf{\Delta}_0  [ \mathbf{\Phi}, \mathbf{G_c} ]$ and thus a functional $\Gamma_0$ of the condensate and connected impurity Green's function:
\begin{eqnarray}
&&\Gamma_0[ \mathbf{\Phi}, \mathbf{G_c} ] = F_\text{imp}[ \mathbf{\Phi}, \mathbf{G_c} ]\nonumber \\
&&+ \int_0^{\beta} d\tau [F_0(\tau)G_c(\tau)+K_0(\tau)\tilde G_c^*(
\tau)+K_0^*(\tau)\tilde G_c(\tau)]\nonumber \\
&&+\frac{1}{N_s\beta}\int_0^\beta d\tau \sum_i [J_0^*(\tau)\phi_i(\tau)+J_0(\tau)\phi_i^*(\tau)],
\end{eqnarray}
where $F_{\text{imp}}$ is the free energy of the local quantum impurity model.
\subsection{Full model}\label{full_model_sec}

The exact functional of the (local) Green's function and condensate are constructed by using the coupling constant integration method, starting from the atomic limit:
\begin{equation}
 \Gamma = \Gamma_{\alpha=1} = \Gamma_0 + \int_0^1 d\alpha \frac{d\Gamma_{\alpha}}{d\alpha}.
 \end{equation}
By using the stationarity of $\Omega$ ($\alpha$-derivatives of the Lagrange multipliers do not contribute) we get
\begin{eqnarray}
\frac{d\Gamma_\alpha}{d\alpha} & = & -\frac{1}{N_s \beta}  \int_0^\beta d\tau t \sum_{\langle i,j \rangle}  \langle  b_i^*(\tau) b_j(\tau) \rangle \nonumber \\ 
&=&-\frac{1}{N_s \beta}  \int_0^\beta d\tau t \sum_{\langle i,j \rangle}  [ \phi_i^*(\tau) \phi_j(\tau) 
+ \langle \delta b_i^*(\tau) \delta b_j(\tau)\rangle  ]\nonumber \\
{} & = & -\frac{1}{2N_s \beta} \text{Tr} \sum_{n, \bf k} \epsilon_{\bf k} \mathbf{G}^{\alpha}_c(\mathbf{k}, i \omega_n) \vert_{ \mathbf{\Phi, G}_c}\nonumber \\
&&-\frac{1}{N_s \beta}  \int_0^\beta d\tau t \sum_{\langle i,j \rangle}  [ \phi_i^*(\tau) \phi_j(\tau) ],\label{dGda}
\end{eqnarray}
where $\omega_n=2\pi n/\beta$ are even Matsubara frequencies and $\epsilon_{\bf k}$ is the dispersion relation of the non-interacting bosons.

We now arrive at the formal expression for the exact functional $\Gamma = \Gamma_{\alpha = 1}$,
\begin{equation}
\Gamma[ \mathbf{\Phi},  \mathbf{G}_c] = \Gamma_0[ \mathbf{\Phi},  \mathbf{G}_c] + \mathcal{K}[ \mathbf{\Phi},  \mathbf{G}_c],
\end{equation}
with the kinetic energy functional given by
\begin{equation}
\mathcal{K}[ \mathbf{\Phi},  \mathbf{G}_c]=\int_0^1 d\alpha \frac{d\Gamma_{\alpha}}{d\alpha}[ \mathbf{\Phi},  \mathbf{G}_c].
\end{equation} 
Requiring stationarity ($\delta \Gamma / \delta \phi_i^*(\tau)=0 $, $\delta \Gamma / \delta \phi_j(\tau)=0 $) determines the value of the source field conjugate to the condensate (assume a homogeneous condensate over the lattice), 
\begin{equation}
\mathbf{J}_0 = zt \mathbf{\Phi}.
\end{equation}
Since the condensate is time-independent (and taken real), we drop the $\tau$ dependence of $J_0$ as well.
The other  stationarity requirement
($\delta \Gamma / \delta G_c=0$, $\delta \Gamma / \delta \tilde G_c=0$) 
determines the hybridization function appearing in the impurity action:
\begin{equation}
F_0(\tau) = \frac{\delta {\mathcal{K}} }{\delta G_c (\tau) }, \hspace{2mm} K_0^*(\tau) = \frac{\delta {\mathcal{K}} }{\delta \tilde G_c (\tau) }.
\end{equation}
Note that for the case $z = \infty$, we have identically $\delta b = 0$ following Appendix A in Ref.\onlinecite{Fisher89} and only static mean-field theory exists, which is physically clear:
For a thermodynamic condensate to be gapless, the $k=0$ component of the Green function should not decay in time. The infinite connectivity of the lattice removes any $k$ dependence and in combination with
bosonic statistics one then sees that the decoupling approximation is exact.
\subsection{Approximation to the kinetic energy functional}

B-DMFT can now be considered as an approximation to the kinetic energy functional.
With the single-particle Green's function of the Bose-Hubbard model in the presence of source fields and for arbitrary coupling constants, we can define a selfenergy
\begin{equation}
\mathbf{G}_c^{\alpha}(\mathbf{k}, i\omega_n) = [ i \omega_n\sigma_3 + (\mu - \alpha \epsilon_{\bf k})\mathbf{I}  + \mathbf{\Delta}_{\alpha} - \mathbf{\Sigma}_{\alpha}[{\bf k}, i \omega_n] ]^{-1},
\end{equation}
where $\sigma_3$ is the Pauli matrix with $\pm 1$ on the diagonal.

The DMFT approximation consists in replacing the selfenergy $\mathbf{\Sigma}_{\alpha}$ for arbitrary $\alpha$ by the impurity model selfenergy $\mathbf{\Sigma}_0$. Hence,
\begin{eqnarray}
&&\mathbf{G}_c^{\alpha}(\mathbf{k}, i\omega_n) \vert_{\text {B-DMFT}}=\nonumber\\ &&\hspace{1cm}[ i \omega_n\sigma_3 + (\mu- \alpha \epsilon_{\bf k})\mathbf{I}  + \mathbf{\Delta}_{\alpha} - \mathbf{\Sigma}_{\alpha = 0}[i \omega_n] ]^{-1}\nonumber \\
&&\hspace{1cm}= [\mathbf{\Delta}_{\alpha}-\mathbf{\Delta}_{0}+\mathbf{G} _c^{-1}-\alpha\epsilon_{\mathbf{k}} \mathbf{I}]^{-1},
\label{G_DMFT}
\end{eqnarray}
where we have used that the impurity selfenergy satisfies the Dyson equation 
\begin{eqnarray}
\mathbf{\Sigma}_{\alpha = 0}[i \omega_n] &=& \mathbf{G}_0^{-1}  - \mathbf{G}_c^{-1} \nonumber\\
&=& i \omega_n\sigma_3 + \mu\mathbf{I} + \mathbf{\Delta}_0 - \mathbf{G}_c^{-1}
\end{eqnarray}
and $\mathbf{G}_0^{-1}$ is the bare Green's function.
Summing over ${\bf k}$, and using the constraint on the local lattice Green's function, we obtain the following relation between $\mathbf{G}_c$ and the hybridization function: 
\begin{eqnarray}
\mathbf{G}_c(i \omega_n) & = & \int d\epsilon D(\epsilon) (\mathbf{\zeta} - \alpha \epsilon\mathbf{I})^{-1} 
= \frac{1}{\alpha} \tilde{D}\Big(\frac{\mathbf{\zeta}}{\alpha} \Big),
\end{eqnarray}
with $\mathbf{\zeta}  =  \mathbf{\Delta}_{\alpha} - \mathbf{\Delta}_{0}  + \mathbf{G}_c^{-1}$.
We used the non-interacting density of states $D(\epsilon) = \frac{1}{N_s} \sum_{\bf k} \delta(\epsilon - \epsilon_{\bf k})$ and its Hilbert-transform $\tilde{D}(\mathbf{z}) = \int d\epsilon D(\epsilon) (\mathbf{z}-\epsilon \mathbf{I})^{-1}$.
By introducing its inverse, $\tilde{D}( R(g)) = g$, the relation above can be inverted ($\alpha R(\alpha \mathbf{G}_c) = \mathbf{\zeta} =  \mathbf{\Delta}_{\alpha} - \mathbf{\Delta}_{0}  + \mathbf{G}_c^{-1}$) and yields the hybridization function as a functional of the local Green's function,
\begin{equation}
\mathbf{\Delta}_{\alpha} [i \omega_n; \mathbf{\Phi}, \mathbf{G}_c] = -\mathbf{G}_c^{-1} +  \mathbf{\Delta}_{0}[\mathbf{\Phi}, \mathbf{G}_c] + \alpha R(\alpha \mathbf{G}_c).
\label{Delta_alpha}
\end{equation}
Inserting the above relation into (\ref{G_DMFT}), the lattice Green's function expressed as a functional of $\mathbf{G}_c$ becomes 
\begin{equation}
\mathbf{G}_c^{\alpha}({\bf k}, i\omega_n) \vert_{\text {B-DMFT}}= (\alpha R(\alpha \mathbf{G}_c) - \alpha \epsilon_{\bf k}\mathbf{I})^{-1}.
\end{equation}
Equation (\ref{dGda}) can now be evaluated with $\mathbf{G}_c^\alpha(\mathbf{k}) \vert_{\text {B-DMFT}}$:
\begin{widetext}
\begin{eqnarray}
&&-\frac{1}{2N_s \beta} \text{Tr} \sum_{n, \bf k} \epsilon_{\bf k} \mathbf{G}_c^{\alpha}({\bf k}, i\omega_n) \vert_{\text {B-DMFT}} -\frac{1}{N_s \beta}  \int_0^\beta d\tau t \sum_{\langle i,j \rangle}  [ \phi_i^*(\tau) \phi_j(\tau) ] \nonumber\\
& &\hspace{10mm}= -\frac{1}{2\alpha \beta} \text{Tr} \sum_n \int d\epsilon \epsilon D(\epsilon) (R(\alpha \mathbf{G}_c )- \epsilon\mathbf{I})^{-1} -\frac{1}{N_s \beta}  \int_0^\beta d\tau t \sum_{\langle i,j \rangle}  [ \phi_i^*(\tau) \phi_j(\tau) ] \nonumber. \nonumber \\
&&\hspace{10mm}= -\frac{1}{2\alpha \beta} \text{Tr} \sum_n \Big[ \alpha \mathbf{G}_c R(\alpha \mathbf{G}_c ) -\mathbf{I} \Big] -\frac{1}{N_s \beta}  \int_0^\beta d\tau t \sum_{\langle i,j \rangle}  [ \phi_i^*(\tau) \phi_j(\tau) ].
\label{dGamma_dalpha}
\end{eqnarray}
\end{widetext}
An explicit expression for the B-DMFT approximation to $\mathcal{K}[\mathbf{\Phi}, \mathbf{G}_c]$ therefore reads
\begin{eqnarray}
\mathcal{K}_{\text{B-DMFT}}[\mathbf{\Phi}, \mathbf{G}_c] &=& -\frac{1}{2\beta} \text{Tr} \sum_n  \int_0^1 d\alpha\left[   \mathbf{G}_c R(\alpha \mathbf{G}_c) - \frac{1}{\alpha}\mathbf{I}  \right] \nonumber \\&-&\frac{1}{N_s \beta}  \int_0^\beta d\tau t \sum_{\langle i,j \rangle}  [ \phi_i^*(\tau) \phi_j(\tau) ],
\label{ekinB-DMFT}
\end{eqnarray}
where the last term reduces to $-zt \phi^* \phi$ for a constant, homogeneous condensate.
 
\subsection{Stationarity conditions}

It immediately follows from Eq.~(\ref{ekinB-DMFT}) that the stationarity condition for the condensate is unaltered in the B-DMFT approximation ($\mathbf{J}_0 = zt \mathbf{\Phi}$), while the stationarity condition for the connected Green's function ($\delta \Gamma / \delta G_c=0$, $\delta \Gamma / \delta \tilde G_c=0$) reads in the B-DMFT approximation (use $R(\alpha G) + \alpha GR'(\alpha G) = \partial_{\rm \alpha}[\alpha R(\alpha G)]$ and the cyclical properties of the trace),
\begin{eqnarray}
&&\mathbf{\Delta}_0[i \omega_n; \mathbf{\Phi}, \mathbf{G}_c] \vert_{\text{B-DMFT}} = -R[ \mathbf{G}_c(i \omega_n) ] + \mathbf{G}_c (i \omega_n)^{-1}\nonumber \\&&\hspace{1cm}= -i\omega_n\sigma_3-\mu\mathbf{I}+\mathbf{\Sigma}_{\text{imp}}+\mathbf{G}_c^{-1}.
\label{hyb}
\end{eqnarray}
Here we have used again the Dyson equation for the second equality.
Applying $\tilde{D}(.)$ to both sides of Eq.~(\ref{hyb}) gives
\begin{equation}
\mathbf{G}_c ( i \omega_n) = \int d\epsilon D(\epsilon)( i \omega_n\sigma_3 + (\mu-\epsilon)\mathbf{I} - \mathbf{\Sigma}_{\rm imp} )^{-1}.
\label{selfconsistency}
\end{equation}
This equation defines the B-DMFT selfconsistency condition.

B-DMFT maps the bosonic lattice problem to a selfconsistent solution of an impurity model, whose action (expressed in terms of the full operators $\bf{b}$, therefore dropping the term $\partial_{\tau}$) now takes the form

\begin{eqnarray}
S_{\text{imp}} & = & - \frac{1}{2}\int_0^{\beta} \int_0^{\beta} d\tau  d\tau' \mathbf{b}^{\dagger}(\tau) {\mathbf{\Delta}}_0 (\tau - \tau') \mathbf{b}(\tau')\nonumber \\ &&-\mu\int_0^\beta d\tau n(\tau) + \frac{U}{2} \int_0^{\beta} d\tau n(\tau) [ n(\tau) - 1]  \nonumber\\
&&  - \mathbf{\Phi}^{\dagger} \Big(zt-\int_0^\beta d\tau'  {\mathbf{\Delta}}_0 (\tau') \Big) \int_0^{\beta} d\tau \mathbf{b}(\tau).
\label{imp_b}
\end{eqnarray}
Assuming $K=K^*$, $\phi=\phi^*$ and dropping all subscripts we can write the action in our final version as

\begin{eqnarray}
S_{\text{imp}} & = & - \frac{1}{2}\int_0^{\beta} \int_0^{\beta} d\tau  d\tau' \mathbf{b}^{\dagger}(\tau) {\mathbf{\Delta}} (\tau - \tau') \mathbf{b}(\tau')\nonumber \\ &&-\mu\int_0^\beta d\tau n(\tau) + \frac{U}{2} \int_0^{\beta} d\tau n(\tau) [ n(\tau) - 1]  \nonumber\\
&&  -\kappa \mathbf{\Phi}^{\dagger} \int_0^{\beta} d\tau \mathbf{b}(\tau)
\label{action}
\end{eqnarray}
where the coefficient $\kappa$ is given by

\begin{equation}
\kappa = zt - \Delta_{11}(i \omega_n = 0) - \Delta_{12}(i \omega_n = 0).
\label{kappa}
\end{equation}

The solution of the impurity problem yields the condensate $\mathbf{\Phi}$ (Eq.~(\ref{phi})), the connected Green's function $\mathbf{G}_c$ (Eq.~(\ref{Gc})) and the selfenergy $\mathbf{\Sigma}_{\text{imp}}$ of the impurity model. The right hand side of Eq.~(\ref{selfconsistency}) then defines the local lattice Green's function, which is identified with the impurity Green's function and thus allows to obtain the new hybridization function for the next iteration by using Eq.~(\ref{hyb}).


\section{DMFT procedure}
\label{sec:DMFT_procedure}

Solving the single site impurity model means computing the local Green's function\begin{equation}
\mathbf{G}(\tau) = -\langle T \textbf{b}(\tau) \textbf{b}^{\dagger}(0) \rangle_{S_\text{imp}}
\label{G}
\end{equation}
and the condensate $\mathbf{\Phi}$ for the impurity action given by Eq.~(\ref{action}).
The hybridization function $\mathbf{\Delta}(\tau)$ and the condensate order parameter $\mathbf{\Phi}$, which is constant in time,
are calculated by a selfconsistent procedure
starting from some arbitrary initial values (usually obtained from the non-interacting or static mean-field limit). We then solve the impurity problem and calculate the new updated parameters $\mathbf{\Delta}(\tau)$ and $\mathbf{\Phi}$ for the action via the Dyson equation. This procedure is repeated until convergence is reached.
The selfconsistency equation for the condensate takes the simple form
\begin{equation}
\mathbf{\Phi} = \langle \mathbf{b}(\tau) \rangle_{S_{\text{imp}}}.
\end{equation}

The new hybridization function $\mathbf{\Delta}(\tau)$ is obtained in the following way:
From Eq.~(\ref{G}) and Eq.~(\ref{phi}) we obtain the connected Green's function via
\begin{equation}
\mathbf{G}_{c}(\tau) = \mathbf{G}(\tau) + \mathbf{\Phi}\mathbf{\Phi}^{\dagger}.
\end{equation}
We then Fourier transform $\mathbf{G}_c(\tau)$ to obtain $\mathbf{G}_c(i\omega_n)$. In Appendix \ref{sec:highf} we show how to obtain accurate Fourier transforms in spite of a finite number of measurement time-steps.
With this we calculate the matrix selfenergy via the Dyson equation
\begin{equation}
\mathbf{\Sigma}(i\omega_n) = \mathbf{G}_0^{-1}(i\omega_n) - \mathbf{G}_{c}^{-1}(i\omega_n).
\end{equation}
Here $\mathbf{G}_0^{-1}(i\omega_n)$ is the bare Green's function which is related to the hybridization function $\mathbf{\Delta}(i\omega_n)$ via
\begin{equation}\label{deltaomega}
\mathbf{\Delta}(i\omega_n) = - i\omega_n \sigma_3 - \tilde{\mu}\mathbf{I} + \mathbf{G}_0^{-1}(i\omega_n).
\end{equation}
The parameter $\tilde{\mu} = \mu - \langle\epsilon\rangle$ is chosen such that
$\mathbf{\Delta}(i \omega_n) \rightarrow 0$ in the limit $\omega_n
\rightarrow \infty$. As we only consider symmetric density of states ($\langle \epsilon \rangle = 0$) here, we write $\mu$ from now on.
Eq.~(\ref{deltaomega}) allows us to rewrite the Dyson equation as
\begin{equation}
\mathbf{\Sigma}(i\omega_n) =  i\omega_n \sigma_3 + \mu \mathbf{I} + \mathbf{\Delta}(i\omega_n) - \mathbf{G}_{c}^{-1}(i\omega_n).
\end{equation}
Employing the DMFT approximation that the lattice selfenergy coincides with the impurity selfenergy, i.e. the selfenergy loses its 
momentum dependence: $\Sigma(\mathbf{k}, i\omega_n)=\Sigma(i\omega_n)$,
we can calculate the $k$-summed (or local) lattice Green's function with
\begin{equation}
\mathbf{G}_\text{latt}(i\omega_n) = \sum_{\mathbf{k}}\Big[i\omega_n\sigma_3+(\mu-\epsilon_{\mathbf{k}})\mathbf{1} -\mathbf{\Sigma}(i\omega_n)\Big]^{-1},
\label{G_lattice_k}
\end{equation}
where $\epsilon_{\mathbf{k}}$ is the dispersion of the lattice. 
For some dispersions $\epsilon_{\mathbf{k}}$ it may be more convenient to transform the summation over wave vectors $\mathbf{k}$ into a integration over the density of states $D(\epsilon)$. With $D(\epsilon) = \sum_{\bf k} \delta(\epsilon - \epsilon_{\bf k})$ Eq.~(\ref{G_lattice_k}) transforms to
\begin{equation}
\mathbf{G}_{\text{latt}} ( i \omega_n) = \int d\epsilon D(\epsilon)[ i \omega_n\sigma_3 + (\mu-\epsilon)\mathbf{I} - \mathbf{\Sigma}(i\omega_n)]^{-1}.
\label{hilbert}
\end{equation}
We now use the Dyson equation again to calculate the new updated hybridization function
\begin{equation}
\mathbf{\Delta}(i\omega_n) = -i\omega_n \sigma_3 - \mu \mathbf{I} + \mathbf{\Sigma}(i\omega_n)  + \mathbf{G}_\text{latt}^{-1}(i\omega_n)
\end{equation}
and the new value for $\kappa$:
\begin{equation}
\kappa = zt - \Delta_{11}(i \omega_n = 0) - \Delta_{12}(i \omega_n = 0).
\end{equation}
After an inverse Fourier transform we obtain $\mathbf{\Delta}(\tau)$ and the selfconsistency loop is closed. We then solve the impurity problem again with the updated action, until convergence is reached.


\subsection{Convergence}
\label{sec:convergence}

In this section we discuss two problems which may occur in the iteration process but can easily be overcome. The first problem is that in the first few iterations, when the solution is still far from the converged result, 
the Hilbert transformation may diverge for some Matsubara frequencies. 
The origin of the problem can be understood by writing Eq.~(\ref{hilbert}) for the individual components\begin{eqnarray}
G_\text{latt}(i\omega_n) &=& \int d\epsilon D(\epsilon) \frac{\zeta^{*} - \epsilon}{|\zeta - \epsilon|^2 - \tilde{\Sigma}^2}, \nonumber \\
\tilde{G}_\text{latt}(i\omega_n) &=& \tilde{\Sigma} \int d\epsilon D(\epsilon) \frac{1}{|\zeta - \epsilon|^2 - \tilde{\Sigma}^2},\label{G_latt_components}
\end{eqnarray}
where $\zeta = i\omega_n + \mu - \Sigma(i\omega_n)$, $\Sigma$ ($\tilde{\Sigma}$) are the diagonal (off-diagonal) components of the matrix selfenergy and we have used that $\tilde{\Sigma}=\tilde{\Sigma}^{*}$.
Obviously, the denominator of Eq.~(\ref{G_latt_components}) can vanish for certain values of  $\Sigma$ and $\tilde{\Sigma}$. 
If the divergence is due to the statistical errors on $\Sigma$ and $\tilde{\Sigma}$, the problem can be cured by running more accurate simulations.
However, another reason for a divergent integral can be that the initial approximations for $\mathbf{\Delta}(\tau)$ and $\mathbf{\Phi}$ are unphysical (``too far away" from the converged solution). 
In almost all cases this problem can be avoided by choosing suitable initial hybridization functions $\mathbf{\Delta}(\tau)$ and condensates $\mathbf{\Phi}$. 
We found that a good choice for the initial $\mathbf{\Phi}$ was the static mean-field value, corresponding to $\mathbf{\Delta}(\tau)=0$. For this value of $\mathbf{\Phi}$ we then calculated $\mathbf{G}_c(\tau)$ and obtained our initial hybridization function $\mathbf{\Delta}(\tau)$ using the lattice Hilbert transform.
Should the integral still be divergent, which only happens in the first few iterations, one can shift (increase) the value of $\Sigma$ such that the integral becomes well defined. Another strategy to fix this problem is to apply a damping procedure to the selfenergy in the first few iterations.

The second problem is related to the convergence properties of the condensate $\mathbf{\Phi}$. Sufficiently far away from the SF-Mott transition we typically need about $10$ to $20$ iterations to reach convergence. Close to the SF-Mott transition, however, the convergence of  $\mathbf{\Phi}$ is very slow, and sometimes up to $500$ iterations are needed to reach a converged solution. With a runtime of a few minutes per iteration this translates into a large numerical effort. Such slowing down of the convergence close to the phase transition can be overcome by using overrelaxation of the condensate $\mathbf{\Phi}$. The updating for $\mathbf{\Phi}$ is changed from Eq.~(\ref{phi}) to
\begin{equation}\mathbf{\Phi} = \alpha (\langle \mathbf{b}(\tau) \rangle_{S_{\text{imp}}} - \mathbf{\Phi}_{\text{old}}) + \mathbf{\Phi}_{\text{old}},
\end{equation}
where $\mathbf{\Phi}_{\text{old}}$ is the condensate from the previous iteration and $\alpha > 1$.


\section{Monte Carlo Method}
\label{solver_sec}

\begin{figure}[h!]
\begin{center}
\includegraphics[angle=0, width=0.8\columnwidth]{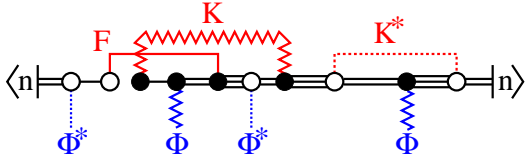}
\caption{(Color online) Diagram corresponding to perturbation orders $m_F=1$, $m_K=1$, $m_{K^*}=1$, $m_\phi=2$, $m_{\phi^*}=2$ and $n(\tau=0)=2$. The hybridization function $F$ determines the amplitude for transitions of bosons from the impurity into the normal reservoir, while operators coupling to source fields $\phi$ and $\phi^*$ represent transitions between the impurity and the BEC reservoir. The off-diagonal hybridization functions $K$ and
$K^{*}$, present only in the BEC phase, represent the amplitudes
for creating or annihilating two bosons at different times.}
\label{diagrams}
\end{center}
\end{figure}

To formulate our Monte Carlo Method we start by writing the impurity action Eq.~(\ref{action}) in non-matrix form:
\begin{widetext}
\begin{eqnarray}
S_\text{imp}&=&- \int_{0}^{\beta}d\tau d\tau' \left[b(\tau) F(\tau-\tau') b^{\dagger}(\tau') + b^{\dagger}(\tau) K(\tau-\tau') b^{\dagger}(\tau') + b(\tau) K^{*}(\tau-\tau') b(\tau')\right]\nonumber \\
&&- \mu \int_0^{\beta} d\tau n(\tau)
+ \frac{U}{2} \int_{0}^{\beta}d\tau n(\tau)[n(\tau)-1] - \kappa \int_{0}^{\beta}d\tau [\phi^{*}b(\tau)+\phi b^{\dagger}(\tau)].
\label{action_no_nambu}
\end{eqnarray}

With this action we expand the partition function $Z=\text{Tr} [T e^{-S_\text{imp}}]$ in powers of the hybridization functions $F$, $K$, $K^*$ and the source fields $\phi$ and $\phi^*$. This leads to the series
\begin{eqnarray}
Z&=&\sum_n \sum_{\mathbf{m}}
\int_0^\beta d\tau^F_{1}\ldots \int_{\tau^F_{m_F-1}}^\beta \!\!\!\!\!\!\! d\tau^F_{m_F} 
\int_0^\beta d\tau'^F_{1}\ldots \int_0^\beta d\tau'^F_{m_F} 
\int_0^\beta d\tau^K_{1}\ldots \int_{\tau^K_{m_K-1}}^\beta \!\!\!\!\!\!\! d\tau^K_{m_K} 
\int_0^\beta d\tau'^K_{1}\ldots \int_0^\beta d\tau'^K_{m_K} \nonumber\\
&&\times \int_0^\beta d\tau^{K^*}_{1}\ldots \int_{\tau^{K^*}_{m_{K^*}-1}}^\beta \!\!\!\!\!\!\! d\tau^{K^*}_{m_{K^*}} 
\int_0^\beta d\tau'^{K^*}_{1}\ldots \int_0^\beta d\tau'^{K^*}_{m_{K^*}} 
\int_0^\beta d\tau^\phi_1\ldots \int_{\tau^\phi_{m_\phi-1}}^\beta \!\!\!\!\!\!\! d\tau^\phi_{m_\phi} 
\int_0^\beta d\tau^{\phi^*}_1\ldots \int_{\tau^{\phi^*}_{m_{\phi^*}-1}}^\beta \!\!\!\!\!\!\! d\tau^{\phi^*}_{m_{\phi^\star}}\nonumber\\
&&\times \Big\langle n \Big|T e^{\mu \int_0^\beta d\tau n(\tau)-U/2\int_0^\beta d\tau n(\tau)[n(\tau)-1]}
b(\tau^F_1)b^\dagger(\tau'^F_1)\ldots b(\tau^F_{m_F})b^\dagger(\tau'^F_{m_F})
b(\tau^{K^*}_1)b(\tau'^{K^*}_1)\ldots b(\tau^{K^*}_{m_{K^*}})b(\tau'^{K^*}_{m_{K^*}})\nonumber\\
&&\hspace{1cm}\times b^\dagger(\tau^{K}_1)b^\dagger(\tau'^{K}_1)\ldots b^\dagger(\tau^{K}_{m_{K}})b^\dagger(\tau'^{K}_{m_{K}})
b^\dagger(\tau^\phi_1)\ldots b^\dagger(\tau^\phi_{m^\phi})b(\tau^{\phi^*}_1)\ldots b(\tau^{\phi^*}_{m_{\phi^*}}) \Big| n\Big\rangle\nonumber\\
&&\times F(\tau^F_1-\tau'^F_1)\ldots F(\tau^F_{m_F}-\tau'^F_{m_F})
K(\tau^K_1-\tau'^K_1)\ldots K(\tau^K_{m_K}-\tau'^K_{m_K})\nonumber\\
&&\hspace{1cm}\times K^*(\tau^{K^*}_1-\tau'^{K^*}_1)\ldots K^*(\tau^{K^*}_{m_{K^*}}- \tau'^{K^*}_{m_{K^*}})
\kappa^{m_\phi+m_{\phi^*}}
\phi(\tau^\phi_1)\ldots\phi(\tau^\phi_{m_\phi})
\phi(\tau^{\phi^*}_1)\ldots\phi(\tau^{\phi^*}_{m_{\phi^*}}),
\end{eqnarray}
where $|n\rangle$ denotes the eigenstate with $n$ bosons and $\mathbf{m}=(m_F, m_K, m_{K^*}, m_{\phi}, m_{\phi^*})$. We can now sample individual diagrams with weight
\begin{align}
&w(n;\tau^F_1,\ldots, \tau^F_{m_F},\tau'^F_1,\ldots, \tau'^F_{m_F};
\tau^K_1,\ldots, \tau^K_{m_K},\tau'^K_1,\ldots, \tau'^K_{m_K};
\tau^{K^*}_1,\ldots, \tau^{K^*}_{m_{K^*}},\tau'^{K^*}_1,\ldots, \tau'^{K^*}_{m_{K^*}};
\tau^\phi_1, \ldots,\tau^\phi_{m_\phi}; \tau^{\phi^*}_1,\ldots,\tau^{\phi^*}_{m_{\phi^*}})=\nonumber\\
&\hspace{1cm} \Big\langle n \Big|T e^{\mu \int_0^\beta d\tau n(\tau)-U/2\int_0^\beta d\tau n(\tau)[n(\tau)-1]}
b(\tau^F_1)b^\dagger(\tau'^F_1)\ldots b(\tau^F_{m_F})b^\dagger(\tau'^F_{m_F})
b(\tau^{K^*}_1)b(\tau'^{K^*}_1)\ldots b(\tau^{K^*}_{m_{K^*}})b(\tau'^{K^*}_{m_{K^*}})\nonumber\\
&\hspace{2cm}\times b^\dagger(\tau^{K}_1)b^\dagger(\tilde \tau^{K}_1)\ldots b^\dagger(\tau^{K}_{m_{K}})b^\dagger(\tau'^{K}_{m_{K}})
b^\dagger(\tau^\phi_1)\ldots b^\dagger(\tau^\phi_{m^\phi})b(\tau^{\phi^*}_1)\ldots b(\tau^{\phi^*}_{m_{\phi^*}}) \Big| n\Big\rangle\nonumber\\
&\hspace{1cm}\times F(\tau^F_1-\tau'^F_1)\ldots F(\tau^F_{m_F}-\tau'^F_{m_F})
K(\tau^K_1-\tau'^K_1)\ldots K(\tau^K_{m_K}-\tau'^K_{m_K})\nonumber\\
&\hspace{2cm}\times K^*(\tau^{K^*}_1-\tau'^{K^*}_1)\ldots K^*(\tau^{K^*}_{m_{K^*}}- \tau'^{K^*}_{m_{K^*}})
\kappa^{m_\phi+m_{\phi^*}}
\phi(\tau^\phi_1)\ldots\phi(\tau^\phi_{m_\phi})
\phi(\tau^{\phi^*}_1)\ldots\phi(\tau^{\phi^*}_{m_{\phi^*}}).
\label{weight}
\end{align}
\end{widetext}
These contributions, illustrated in Fig.~\ref{diagrams}, can be represented by a collection of $m_F+2m_{K^*}+m_\phi=m_F+2m_K+m_{\phi^*}$ creation and annihilation operators on the imaginary time interval $[0,\beta)$. Hybridization functions $F$ connect $m_F$ pairs of creation and annihilation operators, off-diagonal hybridization function $K$ ($K^*$) connect $m_K$ ($m_{K^*}$) pairs of creation (annihilation) operators, while $m_\phi$ ($m_{\phi^*}$) creation (annihilation) operators are linked to source fields $\phi$ ($\phi^*$). The configuration is fully specified by additionally giving the occupation $n$ of the impurity at times $\tau=0$, which in combination with the collection of operators determines $n(\tau)$.

\subsection{Updates and detailed balance}\label{updates_sec}

An ergodic sampling of all possible diagrams requires the following updates:
\begin{enumerate}
\item insertion/removal of a pair $b(\tau)F(\tau-\tau')b^\dagger(\tau')$,
\item exchange of the bath type: \\
{\it a)} $b(\tau)F(\tau-\tau')b^\dagger(\tau')\,\,\, \leftrightarrow \, \kappa\phi^*b(\tau)\kappa \phi b^\dagger(\tau')$,\\
{\it b)} $b(\tau)K^*(\tau-\tau')b(\tau')\,\, \leftrightarrow \, \kappa\phi^*b(\tau)\kappa\phi^*b(\tau')$,\\
{\it c)} $b^\dagger(\tau)K(\tau-\tau')b^\dagger(\tau') \leftrightarrow \, \kappa \phi b^\dagger(\tau)\kappa\phi b^\dagger(\tau')$.
\end{enumerate}

\begin{figure}[h!]
\begin{center}
\includegraphics[angle=0, width=0.9\columnwidth]{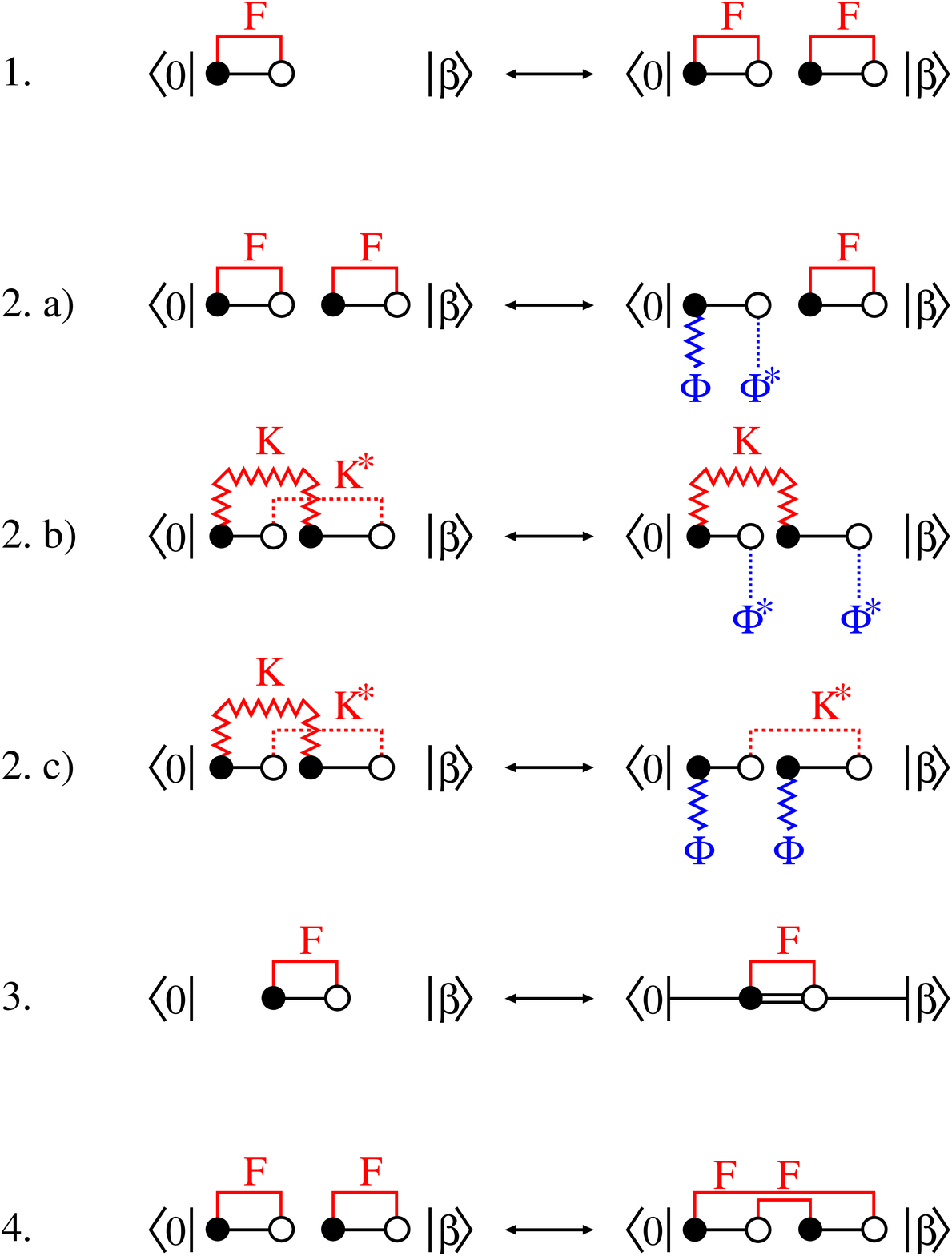}
\caption{(Color online) Illustration of the different updates described in Sec.~\ref{updates_sec}. }
\label{updates}
\end{center}
\end{figure}

In order to improve the efficiency additional moves can/should be used. Updates which may improve the efficiency are shifts of operator positions, moves which reconnect the hybridization lines and moves which increase/decrease $n$ by one. For example, in the normal or Mott phase, where $\phi=K=0$, or close to the SF transition, where $\phi$ is small, it is useful to have an additional move which reconnects two $F$-lines.
The reason for this is that update $1$ is the most time consuming and no $F$-lines can be reconnected via source fields $\phi$. Deep in the Mott phase the insertion of $F$-lines is strongly suppressed. In order to have an efficient sampling, we now need an update which increases/decreases the occupation in the whole imaginary time interval without inserting new operators.
A graphical representation of the updates is shown in Fig.~\ref{updates}.

In order to satisfy detailed balance, we decompose the transition probability to go from state $i$ to state $j$ as
\begin{equation}
p(i \to j) = p^{\text{prob}}(i \to j)\; p^{\text{acc}}(i \to j),
\end{equation}
where $p^{\text{prob}}$ ($p^{\text{acc}}$) is the probability to propose (accept) this move.

The only move which changes the number of operators in the imaginary time interval, and therefore the total perturbation order $m$, is the move to insert or remove an $F$-line.
If we choose to insert an operator pair ($F$-line) at random times $\tau$ and $\tau'$ drawn from a uniform distribution $[0,\beta)$ and propose to remove this pair with probability $1/(m_F+1)$ the proposal probability becomes

\begin{eqnarray}
p^{\text{prob}}(m_F \to m_F+1) &=& \frac{2d\tau d\tau'}{\beta^2}, \\
p^{\text{prob}}(m_F+1 \to m_F) &=& \frac{2}{m_F+1}.
\end{eqnarray}

The factor $2$ comes from the fact that we have to decide if the occupation number is changed in the interval between the two operators (with length $|\tau-\tau'|$) or in the interval which winds around $\tau=\beta$ (with length $\beta-|\tau-\tau'|$). If we choose the interval which winds around $\tau=\beta$ this will change the occupation $n$ at time $\tau=0$.
Denoting the factor $\langle n| \ldots | n\rangle$ in Eq.~(\ref{weight}) by $w_{Tr}(n;\tau^F_1,\ldots, \tau^F_{m_F},\tau'^F_1,\ldots, \tau'^F_{m_F}; \ldots)$, the detailed balance condition for inserting/removing a pair ${b(\tau)F(\tau-\tau')b^\dagger(\tau')}$ becomes
\begin{align}
&\frac{p^{\text{acc}}(m_F\rightarrow m_F+1)}{p^{\text{acc}}(m_F+1 \rightarrow m_F)}=\frac{\beta^2}{m_F+1}F(\tau-\tau')\nonumber\\
&\hspace{5mm}\times \frac{w_{Tr}(n';\tau^F_1,\ldots,\tau,\ldots, \tau^F_{m_F},\tau'^F_1,\ldots, \tau', \ldots \tau'^F_{m_F};\ldots)}
{w_{Tr}(n;\tau^F_1,\ldots, \tau^F_{m_F},\tau'^F_1,\ldots, \tau'^F_{m_F};\ldots)},
\end{align}
where $n'$ and $n$ are different only if the new $F$-line spans over time $\tau = 0$ (and $\tau = \beta$).

For moves which increase/decrease the occupation of the impurity by one we have $p^{\text{prop}}=1$, so the acceptance probabilities satisfy
\begin{align}
& \frac{p^{\text{acc}}(n\rightarrow n+1)}{p^{\text{acc}}(n+1\rightarrow n)} =  \nonumber \\
& \hspace{2mm} \frac{w_{Tr}(n+1;\tau^F_1,\ldots, \tau^F_{m_F},\tau'^F_1,\ldots, \tau'^F_{m_F};\ldots)} {w_{Tr}(n;\tau^F_1,\ldots, \tau^F_{m_F},\tau'^F_1,\ldots, \tau'^F_{m_F};\ldots)}.
\end{align}

Moves which interchange hybridization lines are easiest, since the operator trace term $w_{Tr}$ does not change. The probability to propose such a move is given by the inverse of the number of possibilities to choose the line (or lines) to exchange. For moves which exchange {$b(\tau)F(\tau-\tau')b^\dagger(\tau') \leftrightarrow \kappa \phi^*b(\tau)\kappa\phi b^\dagger(\tau')$} the detailed balance condition reads
\begin{align}
&\frac{p^{\text{acc}}(\{m_F,m_\phi, m_{\phi^*}\}\rightarrow \{m_F-1,m_\phi+1, m_{\phi^*}+1\})}{p^{\text{acc}}(\{m_F-1,m_\phi+1, m_{{\phi^*}}+1\}\rightarrow \{m_F,m_\phi, m_{\phi^*}\})}=\nonumber \\
&\hspace{5mm} \frac{m_F}{(m_\phi+1)(m_{\phi^*}+1)}\frac{\kappa\phi^*\kappa\phi}{F(\tau-\tau')}.
\end{align}
For moves which change an off-diagonal hybridization line into two condensate lines we have to consider that the $K$-lines do not have a defined direction like the $F$-lines. Therefore the probability to choose one $K$-line in a state with $m_K$ $K$-lines is given by $p^{\text{prob}} = 2/m_K$. Similarly, the probability to choose two $\phi$-lines in a state with $m_{\phi}$ $\phi$-lines is given by $p^{\text{prob}} = 2/(m_{\phi} (m_{\phi}-1))$. The detailed balance condition for moves which exchange {$b^{\dagger}(\tau)K(\tau-\tau')b^{\dagger}(\tau') \leftrightarrow \kappa \phi b^{\dagger}(\tau)\kappa\phi b^{\dagger}(\tau')$} is therefore given by
\begin{align}
&\frac{p^{\text{acc}}(\{m_{K}, m_{\phi}\}\rightarrow \{m_{K}-1,m_{\phi}+2\})}{p^{\text{acc}}(\{m_{K}-1, m_{{\phi}}+2\}\rightarrow \{m_{K},m_{\phi}\})}=\nonumber \\
&\hspace{5mm}\frac{m_{K}}{(m_{\phi}+2)(m_{\phi}+1)}\frac{\kappa \phi \kappa \phi}{K(\tau-\tau')},
\end{align}
and similarly for the complex conjugate.

\subsection{Measurement of observables}

By taking functional derivatives of the partition function with respect to either $F(\tau-\tau')$, $K(\tau-\tau')$, $K^{*}(\tau-\tau')$, or both $\phi^{*}(\tau)$ and $\phi(\tau')$ one can calculate the diagonal and off-diagonal parts of the Green's function matrix  $\mathbf{G}(\tau)$, and the condensate order parameter $\phi$. 

For the condensate the functional derivative yields
\begin{equation}
\phi(\tau) = \langle b(\tau) \rangle_{S_{\text{imp}}} = \Big\langle \sum_{i=1}^{m_{\phi^*}} \frac{\delta(\tau,\tau_i^{\phi^*})}{\kappa \phi^{*}(\tau_i^{\phi^*})} \Big\rangle_{MC},
\end{equation}
where $\langle A\rangle_{MC}$ means that the quantity $A$ should be averaged over all configurations obtained in the Monte Carlo sampling. Due to the time independence of $\phi$ this reduces to

\begin{equation}
\phi = \langle b(\tau) \rangle_{S_{\text{imp}}} = \frac{\langle m_{\phi^*}\rangle_{MC} }{\beta \kappa \phi^*}.
\label{phi_meas}
\end{equation}
One can see from Eq.~(\ref{phi_meas}) that one needs a condensate in order to measure a condensate. Therefore one always chooses $\phi \ne 0$ as initial value in the simulation.

The diagonal and off-diagonal Green's function can be measured in two different ways. Either by taking the functional derivative with respect to $F(\tau-\tau')$, $K(\tau-\tau')$ and $K^{*}(\tau-\tau')$ or by differentiating with respect to both $\phi^{*}(\tau)$ and $\phi(\tau')$. This yields

\begin{eqnarray}
G(\tau) &=& -\langle T b(\tau)b^{\dagger}(0)\rangle_{S_{\text{imp}}} \nonumber\\
&=& -\Big\langle \sum_{i=1}^{m_F} \frac{\delta(\tau,
\tau_i^F - \tau_i'^F)}{\beta F(\tau_i^F - \tau_i'^F)} \Big\rangle_{MC} ,\nonumber \\
\tilde{G}(\tau) &=& -\langle T b(\tau)b(0)\rangle \nonumber_{S_{\text{imp}}}  \\
&=& - \Big\langle \sum_{i=1}^{m_{K^*}}
\frac{\delta(\tau, \tau_i^{K^*} - \tau_i'^{K^*})}{\beta K^*(\tau_i^{K^*} -
\tau_i'^{K^*})} \Big\rangle_{MC},
\label{G_meas_hyb}
\end{eqnarray}
in terms of the hybridization and

\begin{eqnarray}
G(\tau) &=& -\langle T b(\tau)b^{\dagger}(0)\rangle_{S_{\text{imp}}} \nonumber\\
&=&- \Big\langle \sum_{i=1}^{m_{\phi^*}} \sum_{j=1}^{m_{\phi}}
\frac{\delta(\tau, \tau_i^{\phi^*} - \tau_j'^{\phi})}{\beta \kappa\phi^{*}
\kappa\phi}\Big\rangle_{MC},\nonumber \\
\tilde{G}(\tau) &=& -\langle T b(\tau)b(0)\rangle \nonumber_{S_{\text{imp}}}  \\
&=& -\Big\langle\sum_{i=1}^{m_{\phi^*}} \sum_{j \neq i=1}^{m_{\phi^*}} \frac{\delta(\tau, \tau_i^{\phi^*} -
\tau_j'^{\phi^*})}{\beta \kappa\phi^{*} \kappa\phi^{*}} \Big\rangle_{MC},
\end{eqnarray}
in terms of the condensate with

\begin{displaymath}
\delta(\tau, \tilde{\tau}) = \left\{ \begin{array}{ll}
\delta(\tau-\tilde{\tau}) & \textrm{for $\tilde{\tau}\geq 0$},\\
\delta(\tau-\tilde{\tau}-\beta) & \textrm{for $\tilde\tau < 0$},
\end{array} \right.
\end{displaymath}
and similarly for the adjoint.
The end point $G(\beta_{-})$ can be accurately measured through the density,
\begin{equation}
G(\beta_{-})=-\left\langle n \right\rangle_{MC} = -\Big\langle \frac{1}{\beta} \int_0^{\beta} d\tau n(\tau) \Big\rangle_{MC}
\end{equation}
and $G(0_{+})=G(\beta_{-}) - 1$. In the condensate phase the expansion order of $\phi$ and $\phi^*$ is much higher than the expansion order of the hybridization, see Sec.~\ref{perturbation_sec}. Therefore, the Green's functions are measured by using the condensate. Close to the SF transition, where $\phi$ is very small, it is more efficient to measure the diagonal Green's function according to Eq.~(\ref{G_meas_hyb}). In the normal or Mott phase, where $\phi=0$, the diagonal Green's function can only be calculated in this way. In practice the off-diagonal Green's function is always measured with $\phi$, even close to the SF transition. This is because the expansion order of $K$ and $K^*$ is always lower than $m_{\phi}$.

The kinetic and potential energy can also be evaluated easily and very accurately. The potential energy is given by

\begin{equation}
E_{\text{pot}} = \frac{U}{2}\langle n(n-1) \rangle - \mu \langle n \rangle,
\end{equation}
and can be computed directly in the Monte Carlo simulation. As shown in Sec.~\ref{full_model_sec} the kinetic energy in the DMFT approximation is given by $E_{\text{kin}} = \left. \frac{d\Gamma}{d\alpha} \right|_{\alpha=1}$. 
Equation~(\ref{dGamma_dalpha}) in the presence of a constant, homogeneous condensate then yields
\begin{equation}
E_{\text{kin}} = -\frac{1}{2\beta} \text{Tr} \sum_n\left[\mathbf{G}_c(\omega_n)R(\mathbf{G}_c(\omega_n))-\mathbf{I} \right] - zt\phi^2.
\end{equation}
By using Eq.~(\ref{Delta_alpha}) and $\mathbf{\Delta}_{\alpha=1}=0$ we obtain
\begin{equation}
E_{\text{kin}} = \frac{1}{2\beta}\text{Tr} \sum_n \left[\mathbf{G}_c(\omega_n)\mathbf{\Delta}_0(\omega_n)\right] - zt\phi^2.
\end{equation}
Going back to imaginary time, we can write this in the individual components as
\begin{eqnarray}
E_{\text{kin}} &=& \int_0^{\beta} d\tau [F(\tau)G_c(\tau) + K(\tau)\tilde G_c^*(\tau) + K^*(\tau)\tilde G_c(\tau)]\nonumber \\
&&- zt\phi^2, 
\end{eqnarray}
or in terms of the full Green's function
\begin{eqnarray}
E_{\text{kin}} &=& \int_0^{\beta} d\tau [F(\tau)G(\tau) + K(\tau)\tilde G^*(\tau) + K^*(\tau)\tilde G(\tau)]\nonumber \\
&&- \kappa\phi^2.
\label{ekin}
\end{eqnarray}
By plugging Eq.~(\ref{phi_meas}) and Eq.~(\ref{G_meas_hyb}) into the above expression we see that the kinetic energy is directly related to the expansion order via
\begin{equation}
E_{\text{kin}} = -\frac{1}{\beta}\langle m_{\text{tot}} \rangle,
\end{equation}
where $m_{\text{tot}}=m_{F}+m_{K}+m_{K^*}+(m_{\phi}+m_{\phi^*})/2$ is the total number of operator pairs in the interval $[0,\beta)$.

This  algorithm is a bosonic version of the hybridization expansion method of Ref.~\onlinecite{Werner06}. Aside from the condensate term and the anomalous hybridization, the main difference lies in the fact that for bosons we don't use any summation of diagrams. In the fermionic case, the analytical summation of all diagrams corresponding to a given sequence of creation and annihilation operators ({\it i.e.,} all diagrams obtained by linking these operators in different ways by hybridization functions) results in a determinant, and was the essential step which allowed to suppress the sign problem and to formulate an efficient algorithm. In a bosonic model, 
a similar summation of diagrams would lead to a matrix permanent whose evaluation is \#P-complete. The only efficient way of evaluating permanents is their stochastic sampling, which is exactly our algorithm of individually sampling the different diagrams (as illustrated in Fig.~\ref{diagrams}) instead of summing them up expicitly. 

Since the off-diagonal hybridization function $K$ is negative, some diagrams have negative weights which leads to a sign problem. However, as we will show later, this becomes an issue only at very low temperatures in the presence of a condensate and does not prevent an accurate computation of phase diagrams and dynamical quantities.

\subsection{Solver test}
To check if the diagrammatic sampling and measurement procedure have been implemented correctly we compare the QMC-result with exact diagonalization. In a Hamiltonian formulation, the impurity model can be written as
\begin{eqnarray}
H_\text{imp}&=&\sum_l \left[V_l (a_l^{\dagger}b +  a_l b^{\dagger}) + W_l (a_l b + a_l^{\dagger}b^{\dagger})\right] + \sum_l \epsilon_l a_l^{\dagger}a_l\nonumber \\
&&- \mu n + \frac{U}{2}n(n-1) -\kappa(\phi^*b+\phi b^{\dagger}).
\end{eqnarray}
The hybridization functions $F$ and $K$ in Eq.~(\ref{action}) are related to the hybridization parameters $V_l$ and $W_l$ through
\begin{eqnarray}
F(i\omega_n)&=&\sum_l \left( \frac{V_lV_l}{\epsilon_l+i\omega_n}+ \frac{W_lW_l}{\epsilon_l-i\omega_n}\right), \\
K(i\omega_n)&=&\frac{1}{2} \sum_l \left( \frac{V_lW_l}{\epsilon_l+i\omega_n}+ \frac{V_lW_l}{\epsilon_l-i\omega_n}\right).
\label{hyb_ED}
\end{eqnarray}
For the test, we represent the bosonic bath by a finite number $n_\text{bath}$ of levels with creation (destruction) operators $a_l^{\dagger}$ ($a_l$) and energies $\epsilon_l$. By Fourier transforming Eq.~(\ref{hyb_ED}) we get
\begin{align}
&\hspace{1.8mm} F(\tau) = \sum_{l=1}^{n_{\text{bath}}} \left( \frac{V_l^2}{e^{\epsilon_l \beta}-1} e^{\epsilon_l \tau} + \frac{W_l^2}{1-e^{-\epsilon_l \beta}} e^{-\epsilon_l \tau} \right), \nonumber \\
&\hspace{1.8mm}K(\tau) = \frac{1}{2} \sum_{l=1}^{n_{\text{bath}}} \left( \frac{V_l W_l}{e^{\epsilon_l \beta}-1} e^{\epsilon_l \tau} + \frac{V_l W_l}{1-e^{-\epsilon_l \beta}} e^{-\epsilon_l \tau}\right).
\end{align}

\begin{figure}[t]
\begin{center}
\includegraphics[angle=0, width=1.0\columnwidth]{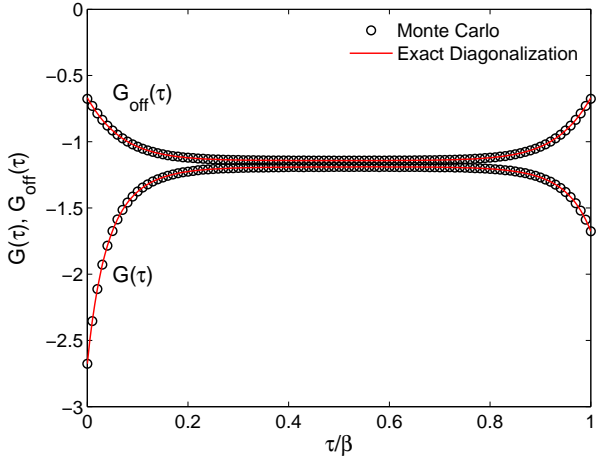}
\caption{(Color online)  Comparison of the diagonal and off-diagonal Green's function calculated by Monte Carlo (black circles) and exact diagonalization (red curve) for a hybridization function which can be described by a bath with one orbital. For parameters, see text.}
\label{comparison_ED_MC}
\end{center}
\end{figure}

For the comparison we choose one orbital ($n_{\text{bath}}=1$) and the following parameters: $t=1$, $\beta =1$, $\mu=20$, $U=20$, $\kappa=6$, $\phi=1$, $\epsilon=1$, $V=1$ and $W=-0.2$. In Fig.~\ref{comparison_ED_MC} we show the diagonal and off-diagonal Green's function calculated with exact diagonalization (ED) and Monte Carlo simulations.

The density and condensate order parameter are $n_{\text{ED}}=1.6762$ and $\phi_{\text{ED}}=1.07976$ for ED, and $\langle n_{\text{MC}} \rangle =1.67619(3)$ and $\langle \phi_{\text{MC}} \rangle = 1.07974(9)$ for the Monte Carlo simulations. Note that for this set of parameters the average sign $\langle s \rangle $ in the Monte Carlo simulation is $\langle s \rangle = 0.40695(7)$. The perfect agreement with the ED result shows that the diagrammatic sampling and measurement procedure have been implemented correctly.
Since the model we considered here contains both diagonal and off-diagonal hybridization terms and since a bath with a finite number of levels is as difficult to treat as any other bath, this serves as a nontrivial test for the Monte Carlo solver.

\subsection{Perturbation order and average sign}\label{perturbation_sec}
In this section we show how the expansion order of the hybridization function scales with interaction and how the sign problem, caused by the off-diagonal hybridization $K$ scales at the Mott insulator  (or normal phase) to SF transition. Since we sample the operator configuration in the imaginary time interval $[0,\beta)$ the perturbation order grows roughly proportional to $\beta$ in all phases.

\begin{figure}[t]
\begin{center}
\includegraphics[angle=0, width=1.0\columnwidth]{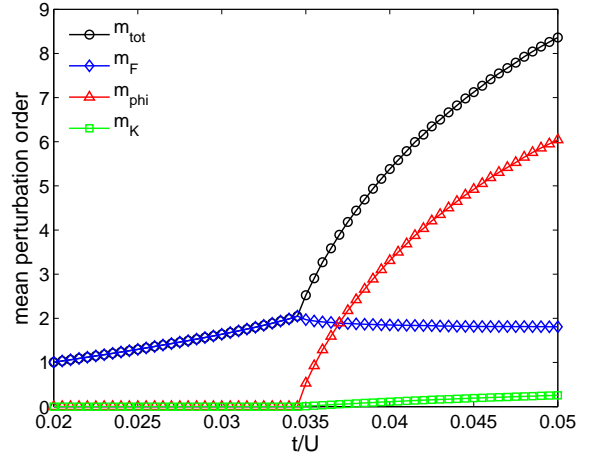}
\caption{(Color online)  Mean perturbation order vs. $t/U$ at filling $n =1$ and inverse temperature $\beta t = 2$ for the 3d cubic lattice. The superfluid phase sets in where $m_{\phi}$ and $m_K$ become non-zero.}
\label{perturbation_vs_U}
\end{center}
\end{figure}

In Fig.~\ref{perturbation_vs_U} we show how the mean perturbation order grows as one goes from the normal phase to the SF phase at filling $n=1$. In the normal phase we have only contributions from the diagonal hybridization $F$ and the perturbation order $m_F$ decreases with increasing $U$. In the SF phase the perturbation order grows rapidly with decreasing interaction $U/t$, mainly because of the condensate contribution $m_{\phi}$ and the off-diagonal hybridization contribution $m_{K}$.

\begin{figure}[t]
\begin{center}
\includegraphics[angle=0, width=1.0\columnwidth]{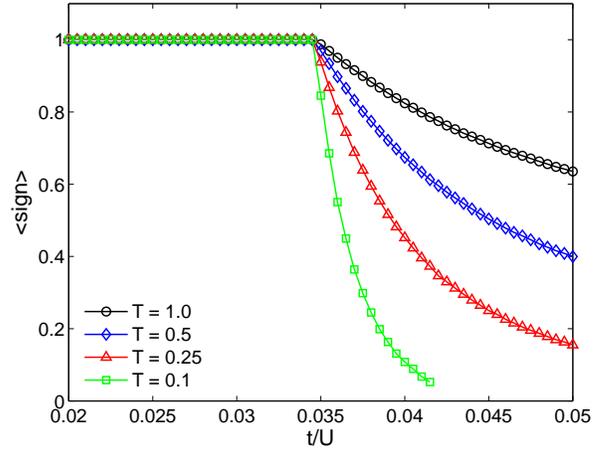}
\caption{(Color online)  Average sign vs. $t/U$ at filling $n =1$ for the 3d cubic lattice at different temperatures. The sign is negative (positive) for odd (even) expansion orders in $m_K$. Hence, where the average sign starts to deviate from 1 is indicative of the phase transition into the superfluid phase.}
\label{sign_vs_U}
\end{center}
\end{figure}

Figure~\ref{sign_vs_U} illustrates the sign problem we encounter when going from the Mott phase to the SF phase. At the phase boundary where the condensate vanishes (and therefore $K$ vanishes) the sign problem disappears. Therefore the sign problem does not prevent an accurate computation of phase diagrams. The sign problem is only severe deep in the condensate phase.

\section{Simple Limits}\label{limits_sec}

\subsection{Non-interacting bosons}

For an ideal Bose gas ($U=0$), the total number of particles is given by
\begin{equation}
n = |\phi|^2 + \int d\epsilon \frac{D(\epsilon)}{e^{\beta(\epsilon-\mu)}-1},  \hspace{1.0cm} |\epsilon| \le zt,
\end{equation}
where $zt$ is the half-bandwidth of the lattice. Just like for an ideal Bose gas in continuous space, the chemical potential $\mu$ has to be lower than the bottom of the band, $\mu \le -zt$ in order to keep the number of particles finite.

For temperatures $T>T_c$, the condensate vanishes, $|\phi|^2=0$ and the above equation determines the density $n$ as a function of the chemical potential $\mu$. For $T<T_c$ the chemical potential must be pinned at the lower edge of the band, $\mu=-zt$ in order to have condensation.  The total number of particles is then a function of the condensate density $|\phi|^2$.
In the non-interacting limit the off-diagonal hybridization functions $K$ and $K^*$ vanish and the B-DMFT equations equations become exact. The impurity action now takes the simple form
\begin{eqnarray}
S_\text{imp}&=&-\int_{0}^{\beta}d\tau d\tau'  b(\tau) F(\tau-\tau')
b^{\dagger}(\tau') - \int_{0}^{\beta}d\tau n(\tau)\nonumber \\
&&- \kappa \int_{0}^{\beta}d\tau
[\phi^{*}(\tau)b(\tau)+\phi(\tau)b^{\dagger}(\tau)].
\end{eqnarray}

This is a quadratic action that can be solved analytically. The solution for the non-interacting Green's function and hybridization function is given by
\begin{eqnarray}
G(i\omega_n) &=& G_0(i\omega_n) = \int d\epsilon
\frac{D(\epsilon)}{i\omega_n + \mu - \epsilon},  \hspace{1.0cm} (|\epsilon| \le zt) \nonumber \\
F(i\omega_n) &=& i\omega_n - \mu +G_0^{-1}(-i\omega_n).
\label{G_U0}
\end{eqnarray}

\subsection{Static mean-field}

One obtains a selfconsistent mean-field theory (the decoupling approximation) by substituting
\begin{eqnarray}
b_i^{\dagger} b_j &=& \langle b_i^{\dagger} \rangle b_j + b_i^{\dagger} \langle b_j \rangle - \langle b_i^{\dagger} \rangle \langle b_j \rangle \nonumber \\
&=& \phi (b_i^{\dagger} + b_j) - \phi^2
\end{eqnarray}
into our Hamiltonian defined by Eq.~(\ref{hamiltonian}). If one drops the term $-\phi^2$, which is just a constant shift in energy one obtains the following Hamiltonian
\begin{equation}
H_{\text{MF}} = -\kappa \phi \sum_i (b_i^{\dagger} + b_i) + \frac{U}{2} \sum_i n_i(n_i-1) - \mu \sum_i n_i.
\end{equation}
where $\kappa = \sum_{\langle \cdot, j \rangle} t = zt$ is the hopping term summed over the nearest neighbors. For the $3d$ cubic lattice ($z=6$) this just gives the half-bandwidth $\kappa = 6t$. This Hamiltonian can be expressed as a matrix in the occupation number basis (truncated at some maximum occupation) and solved by exact diagonalization. One chooses an initial value for the condensate $\phi$ and determines $\phi$ iteratively by solving $\phi = \langle b \rangle$ until convergence is reached.

By using B-DMFT we can reproduce the static mean-field results by setting the hybridization function to zero, i.e. $\mathbf{\Delta}(\tau) = 0$. Since there is no hybridization Eq.~(\ref{kappa}) reduces to $\kappa=zt$. 

\section{Numerical results}
\label{results_sec}

In this section we present results for the Bose-Hubbard model on a 3d cubic and 2d square lattice obtained with B-DMFT. All our results are compared to other methods like static mean-field theory, worm-type quantum Monte Carlo (QMC) simulation on a lattice of up to $40^3$ sites~\cite{Prokofev98, Pollet07_LOWA} and with a recently developed numerically exact method on the Bethe lattice.~\cite{Semerjian09} Since the Bethe lattice can be also directly simulated with B-DMFT we will show a direct comparison between B-DMFT and the numerically exact solution.

\begin{figure}[ht]
\begin{center}
\includegraphics[angle=0, width=1.0\columnwidth]{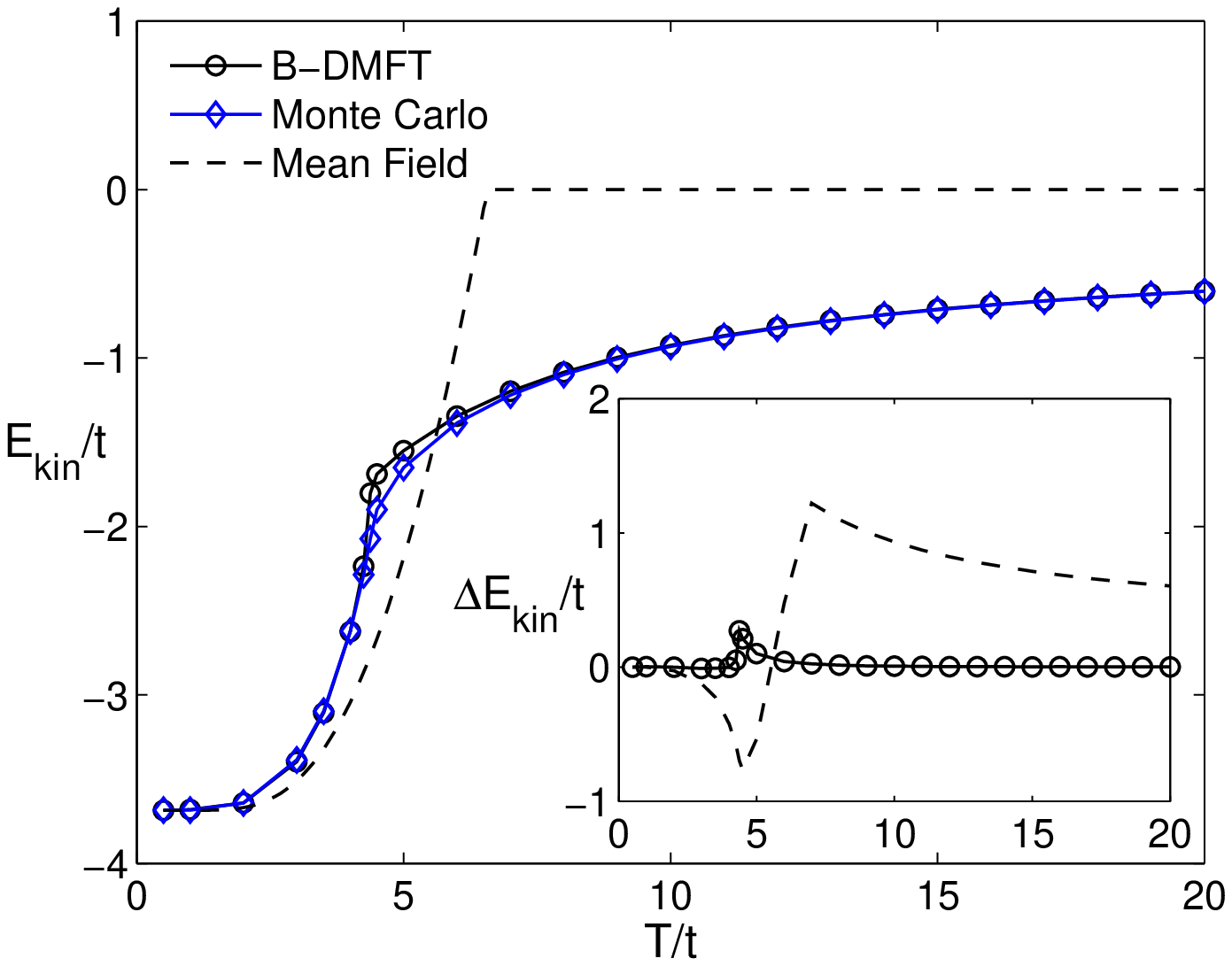}
\caption{(Color online)  Kinetic energy of the Bose-Hubbard model on the 3d cubic lattice as a function of temperature for $\mu/t=8$ and $U/t=20$ ($n\approx1$). Results obtained from B-DMFT (black circles) are compared to lattice QMC (blue diamonds) and to static mean-field theory (black dashed line). Inset: Energy difference from the QMC data for the same parameters. The QMC results are obtained on a lattice with $10^3$ sites except close to the transition ($4\le T/t \le6$) where $40^3$ sites were used. Error bars are smaller than the symbol size.}
\label{figure_ekin}
\end{center}
\end{figure}

\begin{figure}[ht]
\begin{center}
\includegraphics[angle=0, width=1.0\columnwidth]{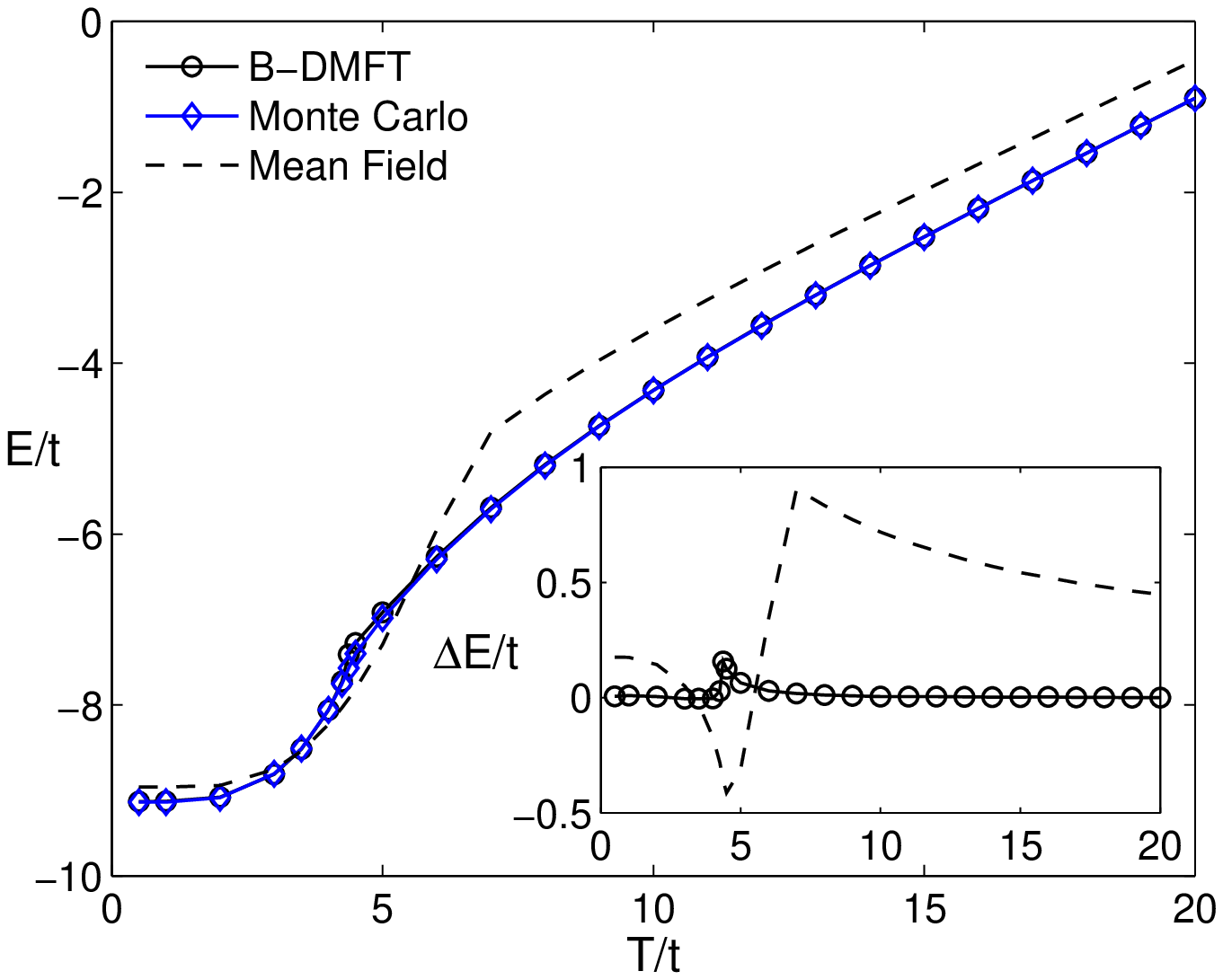}
\caption{(Color online)  Total energy of the Bose-Hubbard model on the 3d cubic lattice as a function of temperature for $\mu/t=8$ and $U/t=20$ ($n\approx1$). Results obtained from B-DMFT (black circles) are compared to lattice QMC (blue diamonds) and to static mean-field theory (black dashed line). Inset: Energy difference from the QMC data for the same parameters. The QMC results are obtained on a lattice with $10^3$ sites except close to the transition ($4\le T/t \le6$) where $40^3$ sites were used. Error bars are smaller than the symbol size.}
\label{figure_etot}
\end{center}
\end{figure}

\subsection{3d cubic lattice} \label{3d_sec}

\subsubsection{Kinetic and total energy}

In this section we present the results for the 3d cubic lattice. Since DMFT can be considered as an approximation to the kinetic energy functional, Eq.~(\ref{ekinB-DMFT}), it is interesting to see how the kinetic energy per site obtained from B-DMFT compares to results from lattice Monte Carlo simulations~\cite{CapogrossoSansone07} and static mean-field theory. In Fig.~\ref{figure_ekin} we show the kinetic energy as a function of temperature as one goes from the SF to the normal phase. In the case of static mean-field theory, the kinetic energy is just given by $E_{\text{kin}} = -zt\phi^2$, where $\phi=\langle b \rangle$ is the condensate order parameter. In the ground state regime this gives a good approximation of the kinetic energy. For $T>T_c$, where $\phi=0$, the kinetic energy vanishes since hopping of the normal bosons is completely neglected. The agreement of the B-DMFT results with the exact QMC results is excellent over the whole temperature range. Only close to $T_c$ there is some small deviation. 
For the QMC simulation we used $10^3$ lattice sites, except for temperatures in the range $4\le T/t \le6$ where $40^3$ sites where used. In Fig.~\ref{figure_etot} we show the total energy as a function of temperature. The remarkable accuracy of the B-DMFT result for the total energy at all temperatures implies that entropies may also be computed reliably using B-DMFT.

\begin{figure}[ht]
\begin{center}
\includegraphics[angle=0, width=1.0\columnwidth]{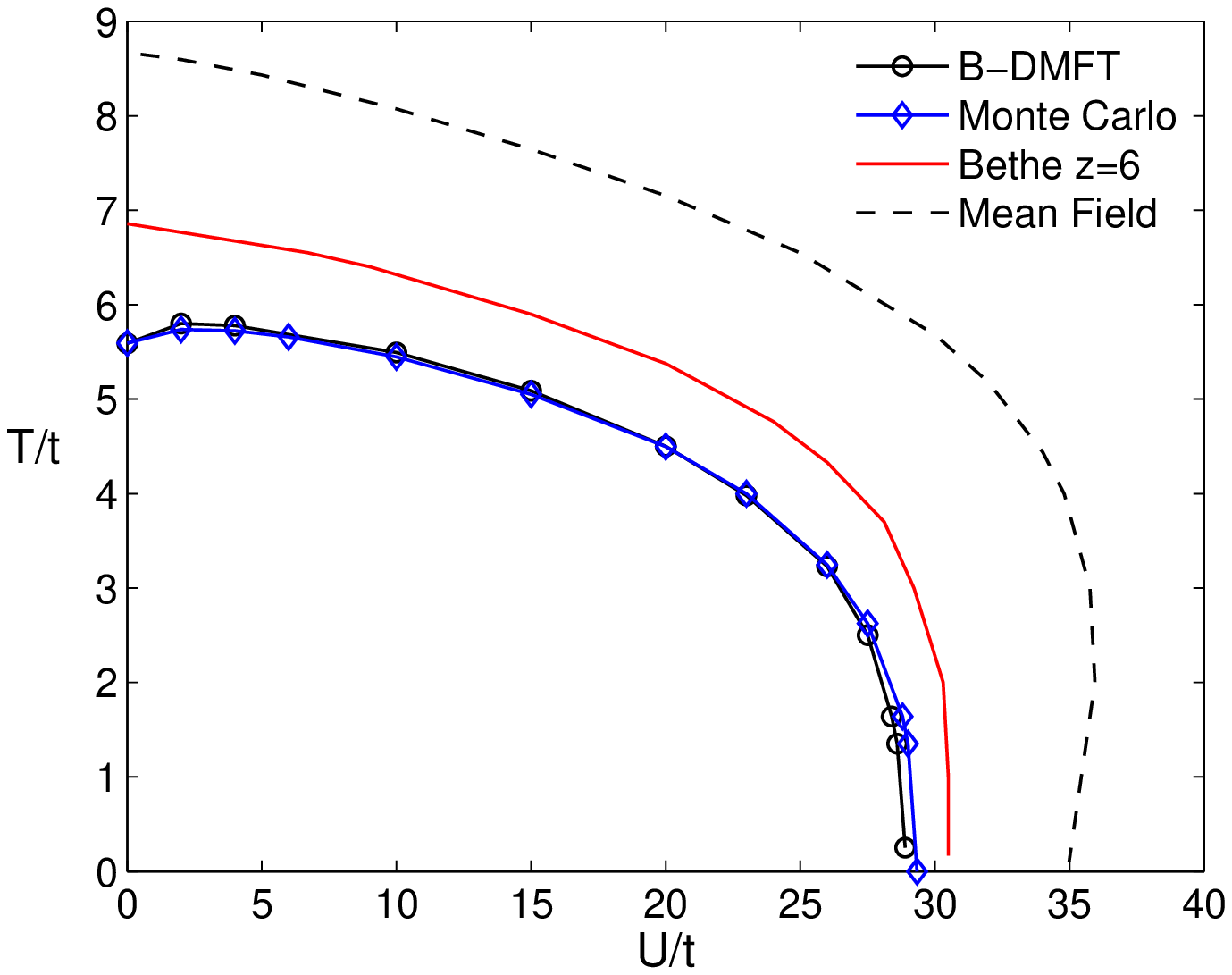}
\includegraphics[angle=0, width=1.0\columnwidth]{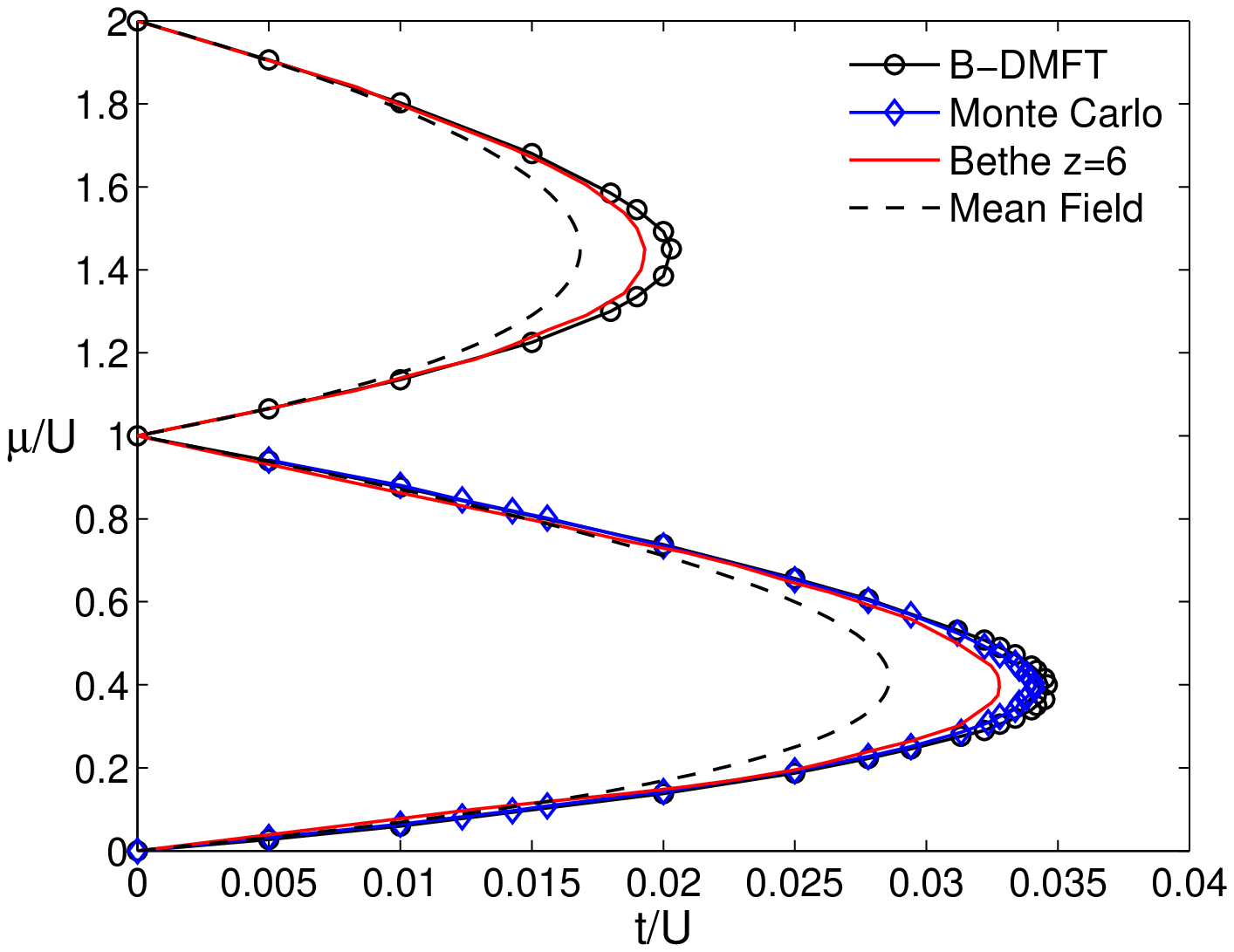}
\caption{(Color online)  Top panel: Phase diagram (superfluid to normal liquid transition) of the cubic lattice Bose-Hubbard model in the space of interaction and temperature for $n=1$. The dashed line shows the static mean-field result, the red curve the exact solution for a Bethe lattice with coordination number $z=6$ (Ref.~\onlinecite{Semerjian09}) and the blue curve with open diamonds the QMC result from lattice simulations (Ref.~\onlinecite{CapogrossoSansone07}). The black line with open circles corresponds to the B-DMFT solution, which yields a second order transition.
Bottom panel: ground-state phase diagram in the space of $t/U$ and $\mu/U$, showing the first two Mott lobes surrounded by superfluid. The B-DMFT phase boundary was computed at $\beta t=2$. Error bars are much smaller than the symbol size.}
\label{ground_state_3d}
\end{center}
\end{figure}

In Fig.~\ref{ground_state_3d} we show the finite temperature phase diagram (top panel) and the ground state phase diagram (bottom panel) for the first and second Mott lobe of the Bose-Hubbard model on a 3d cubic lattice and compare results obtained with B-DMFT to exact results from lattice QMC simulations, the exact solution for the Bethe lattice with coordination number $z=6$,~\cite{Semerjian09} and to static mean-field results.\cite{Anders10}
The SF phase is characterized by a finite value of $\phi=\langle b \rangle$, while we have $\phi=\langle b \rangle=0$ in the Mott insulating and normal phase. For the calculation of the ground state phase diagram we used $\beta t=2$, which is shown in Fig.~\ref{ground_state_3d} to be a sufficiently low temperature. Based on simulations done at $\beta t=4$ and $\beta t=8$ we found that  any systematic error is smaller than the statistical error.
The excellent agreement between our B-DMFT results and the full solution of the Bose-Hubbard model shows that the Mott-transition is a local phenomenon, well described by a momentum-independent selfenergy and that the condensed bosons are accurately described by a uniform condensate.

\subsubsection{Critical exponents}
\label{sec:critical_exponents}

We now examine the critical exponent of the order parameter $\phi$. In Ref.~\onlinecite{Koga10} fermionic DMFT was employed to study the superfluid state. The temperature dependence of the condensate order parameter $\Delta$ goes as $\Delta \sim |(T_c-T)/T_c|^{\beta}$ in the vicinity of the critical temperature $T_c$, with $\beta=1/2$ being the mean-field exponent. The symmetry breaking happens on the mean-field level in the two-particle channel given by a spin-up and a spin-down fermion. For any fermionic operator we have that $\langle c \rangle \equiv 0$ such that condensation can occur the earliest in the two-particle channel, {\it e.g.,} $ \langle c^{\dagger}_{\uparrow} c_{\downarrow} \rangle$ or $ \langle c_{\uparrow}c_{\downarrow} \rangle $ may be non-zero. For a bosonic operator, a single operator $b$ may already have a non-zero expectation value, $\langle b \rangle \neq 0$, leading to condensation at the one-body level. The B-DMFT action allows for this, while at the two-body level ({\it i.e.,} the selfconsistently determined hybridization terms) the one-loop correction to the condensate will change the critical exponents from their mean-field values.  The exponents we obtain are therefore not universal but depend on the parameters $t$, $\mu$, $U$, $\beta$ and the lattice dispersion $\epsilon_{\mathbf{k}}$. Whenever $\kappa=\text{const}$, we recover the mean-field exponents. This happens of course in the static mean-field limit $(\kappa=zt)$ and for non-interacting bosons $(\kappa=-G_0^{-1}(i\omega_n=0))$, where the mean-field exponent $\beta=1/2$ is exact. In Ref.~\onlinecite{Semerjian09} mean-field exponents were also found for the exact solution on the Bethe lattice.


\begin{figure}[t]
\begin{center}
\includegraphics[angle=0, width=1.0\columnwidth]{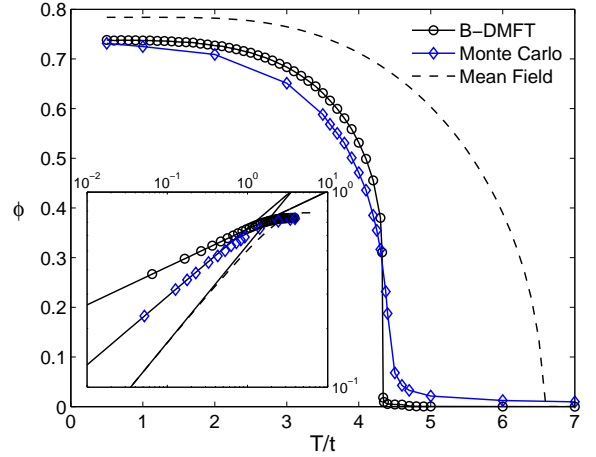}
\caption{(Color online)  Main panel: Condensate order parameter $\phi$ as a function of temperature of the Bose-Hubbard model on the 3d cubic lattice for $\mu/t=8$ and $U/t=20$ ($n\approx1$). Results obtained from B-DMFT (black circles) are compared to lattice QMC (blue diamonds) and to static mean-field theory (black dashed line). Inset:  Zooming in on the region close to the SF-normal transition showing the critical exponent $\beta$. Here $\phi$ is plotted as a function of $T_c-T$ for $T<T_c$. The full lines are fits to $\phi=A|T_c-T|^\beta$ (B-DMFT: $\beta=0.194(2)$, $T_c/t=4.365(3)$, $At=0.6463$, QMC: $\beta=0.35$, $T_c/t=4.43(3)$, $At=0.6517$, Mean-field: $\beta=0.5$, $T_c/t=6.5661$, $At=0.535$).
The QMC results are obtained on a lattice with $40^3$. Error bars are smaller than the symbol size.}
\label{figure_exponents_3d}
\end{center}
\end{figure}

In Fig.~\ref{figure_exponents_3d} (main panel) we show the condensate order parameter $\phi$ as a function of temperature obtained by B-DMFT and compare the results to lattice QMC and static mean-field theory. Close to the critical point the condensate goes as $\phi \sim |(T_c-T)/T_c|^{\beta}$ for $T<T_c$. The QMC results are obtained on a lattice with $40^3$ sites, which still leads to a rounding of the data compared to the thermodynamic limit. They are based on the $k=0$ component of the Fourier transform of the equal-time density matrix.
By plotting $\phi$ as a function of $T_c-T$ in a log-log plot we can extract the critical exponent $\beta$ from the slope of the line as $T \to T_c$, which is shown in the inset of Fig.~\ref{figure_exponents_3d}. Static mean-field theory of course gives the mean-field exponent $\beta=1/2$. The exact model belongs to the 3d XY universality class with the exponent $\beta \approx 0.35$,~\cite{Campostrini06} while the exponent obtained from B-DMFT for these parameters is $\beta \approx 0.19$.

\subsubsection{The weakly interacting Bose gas regime}
\label{sec:Hugenholtz_Pines}

When bosonic field theories are expanded in $U$ (cf. the weakly interacting Bose gas of Ref.~\onlinecite{CapogrossoSansone10}), the effect of the chemical potential is non-perturbative. The chemical potential is negative for the ideal Bose gas, and has to change sign in the presence of repulsive interactions. 
 There is an implicit relation between the condensate density $n_0 = |\phi|^2$ and the chemical potential, which follows from the condition of thermodynamic equilibrium,\cite{FetterWalecka}
\begin{equation}
\left[ \frac{\partial \Omega(T=0, V, \mu, n_0 )}{ \partial n_0} \right]_{V, \mu} = 0,
\end{equation}
leading to $\mu = \langle 0 \vert \frac{\partial \hat{H}_{\rm int}}{\partial n_0}  \vert 0 \rangle$, with $\hat{H}_{\rm int}$ the interacting two-body terms and  $ \vert 0 \rangle$ the ground state (finite temperature extensions also exist\cite{CapogrossoSansone10}). In any expansion order this remains valid and can be worked out to yield
\begin{equation}
\mu = \Sigma(k=0, \omega = 0) - \tilde{\Sigma}(k=0, \omega = 0),
\end{equation}
which is the famous relation for a gapless spectrum first derived by Hugenholtz and Pines.~\cite{HugenholtzPines}

\begin{figure}[t]
\begin{center}
\includegraphics[angle=0, width=1.0\columnwidth]{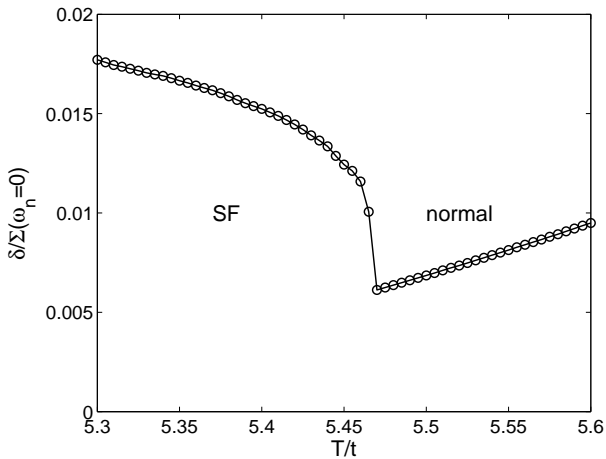}
\caption{Relative deviation from the Hugenholtz-Pines relation, $\delta/\Sigma(\omega_n=0)$ with $\delta=\Sigma(\omega_n=0)-\tilde\Sigma(\omega_n=0)-\mu-zt$, of the Bose-Hubbard model on the 3d cubic lattice as a function of temperature. The figure shows the transition from the SF to the normal phase for $U/t=10$ at filling $n\approx 1$. Error bars are of the order of the symbol size.}
\label{figure_hugenholtz}
\end{center}
\end{figure}

The Hugenholtz-Pines relation does not hold in our B-DMFT formalism. To start the discussion, let us consider a mean-field approach in which the condensate is determined selfconsistently but the hybridization functions are not. Then it is easy to see that the Hugenholtz-Pines relation is shifted to $\mu = \Sigma(\omega = 0) - \tilde{\Sigma}(\omega = 0) - zt$. If, however, the hybridization functions are also determined selfconsistently (as in the full B-DMFT scheme), then they will also depend on and influence the condensate density (and contain $\mu$ in them), leading to the break-down of the Hugenholtz-Pines relation. Indeed, although Ref.~\onlinecite{CapogrossoSansone10} uses a skeleton approach for the description of the WIBG, the discussion of the Hugenholtz-Pines relation is provided in terms of bare Green functions in order not to mix the expansion orders.

However, it turns out that the deviations from the (shifted) Hugenholtz-Pines theorem are minimal. This is shown in Fig.~\ref{figure_hugenholtz} where we plot the relative deviation from the Hugenholtz-Pines relation at a value of $U/t=10$ 
across the SF to normal transition. One clearly sees that the deviation grows as one moves deeper into the SF phase. In the Mott and normal phases, where $\phi=0$, the Hugenholtz-Pines relation is not valid.

\begin{figure}[t]
\begin{center}
\includegraphics[angle=0, width=1.0\columnwidth]{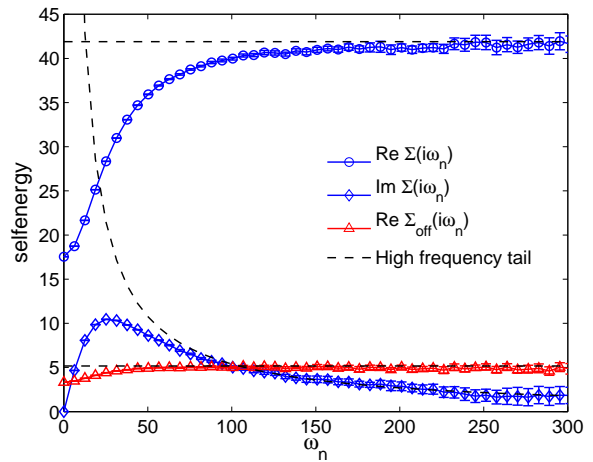}
\caption{(Color online)  Selfenergy of the Bose-Hubbard model on the 3d cubic lattice as a function of Matsubara frequencies in the SF phase, $\beta t=1$, $\mu/t=8$ and $U/t=20$ ($ \langle n \rangle \approx1$). The curve with blue circles (diamonds) shows the real (imaginary) part of the selfenergy $\Sigma(i\omega_n)$ and the curve with red triangles shows the off-diagonal selfenergy $\tilde \Sigma(i\omega_n)$. The dashed lines show the analytically known high frequency tail where $\Sigma(i\omega_n)=\Sigma_0+\Sigma_1/i\omega_n+O(1/(i\omega_n)^2)$ and $\tilde\Sigma(i\omega_n)=\tilde\Sigma_0+O(1/(i\omega_n)^2)$ with $\Sigma_0=2U\langle n \rangle$, $\Sigma_1=U^2\left(3\langle n^2 \rangle - \langle n \rangle - 4\langle n \rangle^2 + \langle bb\rangle^2\right)$ and $\tilde \Sigma_0 = U\langle bb \rangle$.}
\label{figure_selfenergy}
\end{center}
\end{figure}

As already mentioned in the introduction, it was shown in Refs.~\onlinecite{Nepomnyashchii78, Nepomnyashchii83} that gapless field theories have selfenergies that should be momentum dependent and, in particular, that the anomalous selfenergy has to go to zero for small momenta at zero frequency. 
In Beliaev's diagrammatic approach of the weakly interacting Bose gas, the first order selfenergies are momentum and frequency independent, and this condition is not satisfied, questioning Beliaev's approach. Similarly, in the B-DMFT scheme the selfenergies are momentum independent.  In Fig.~\ref{figure_selfenergy} one sees that the anomalous selfenergy is non-zero for zero frequency, as could be expected from the explicit symmetry breaking terms in the B-DMFT action. On the other hand, the arguments of Ref.~\onlinecite{CapogrossoSansone10} for the weakly interacting Bose gas also hold for B-DMFT. In particular, the authors of Ref.~\onlinecite{CapogrossoSansone10} argued that the system reaches its thermodynamic limit for properties such as the energy and the entropy for small system sizes. 
In Figs.~\ref{figure_ekin}, \ref{figure_etot} we have already seen that the (kinetic) energy is reproduced remarkably well over the entire temperature range in 3d, except in the fluctuation region near the normal-superfluid phase transition point.

From the viewpoint that DMFT and B-DMFT sum up all skeleton diagrams built with local propagators only (which is an (asymptotic) expansion in $U$), we expect that our B-DMFT theory should recover the physics of the weakly interacting Bose gas (Bogoliubov Hamiltonian), which is known to be an excellent approximation at low values of $nU$. It is the only limit for which it is not obvious that the B-DMFT formalism will be successful, while the other regimes of strong $U$ or high temperature are well captured by a single-site action.
In the skeleton approach to the WIBG theory,~\cite{CapogrossoSansone10} the selfenergies are to leading and subleading order given by  $\Sigma = 2 \langle n \rangle U$ and $\tilde\Sigma = \langle bb \rangle U$ (see Fig.~\ref{fig:WIBG_diagram}). 
To this order of accuracy, the chemical potential is given by  $\mu^{(1)} = 2\langle n \rangle U - \langle bb \rangle U \approx \langle n_0 \rangle U$. 
The latter approximation is made to keep the equations simple (it is not fundamental) since we don't want to go into technical details of the WIBG theory here~\cite{CapogrossoSansone10}
but it allows us to simply relate the condensate $\langle n_0 \rangle $ to the chemical potential $\mu^{(1)} \approx \mu$ in this limit, after also taking the shift $-6t$ from the band edge into account. Because the chemical potential in B-DMFT is an input parameter and fully renormalized from the start, this is another way of understanding the deviations from the Hugenholtz-Pines relation.

\begin{figure}[t]
\begin{center}
\includegraphics[angle=0, width=1.0\columnwidth]{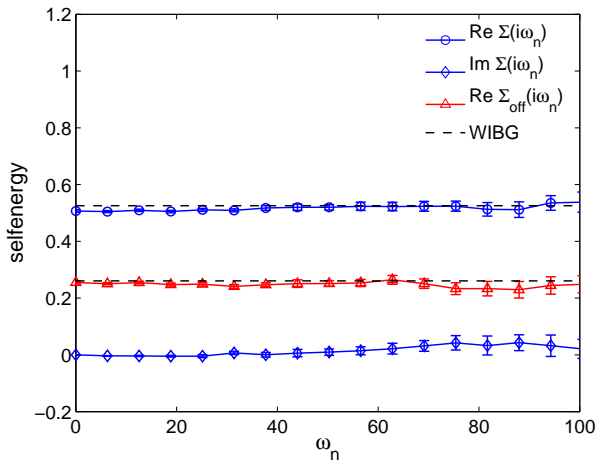}
\caption{(Color online)  Selfenergy of the Bose-Hubbard model on the 3d cubic lattice as a function of Matsubara frequencies in the SF phase, $\beta t=4$, $\mu/t=-5.75$ and $U/t=0.5$ ($ \langle n \rangle \approx0.5$). The parameters are in the regime where the weakly interacting Bose gas (WIBG) theory applies, which predicts $\Sigma^{(1)} = 2 \langle n \rangle U$ and $\tilde\Sigma^{(1)} = \langle bb \rangle U$. The selfenergies are almost frequency independent and are in very good agreement with the WIBG predictions.}
\label{figure_selfenergy_WIBG_3d}
\end{center}
\end{figure}
 
We now discuss the numerical results for the weakly interacting Bose gas limit. 
In order to see some depletion effects, we choose parameters such that $ \langle n \rangle =0.5$ and $\beta t = 4$. The main result can be seen in Fig.~\ref{figure_selfenergy_WIBG_3d}, where one sees that the selfenergy and the anomalous selfenergy become frequency independent and approach their WIBG values, which are precisely the leading terms in the high-frequency expansions used in B-DMFT: 
For the normal selfenergy it is $\Sigma = 2 \langle n \rangle U$, while for the anomalous selfenergy it is $U\langle b b \rangle$.

\begin{figure}[t]
\begin{center}
\includegraphics[angle=0, width=1.0\columnwidth]{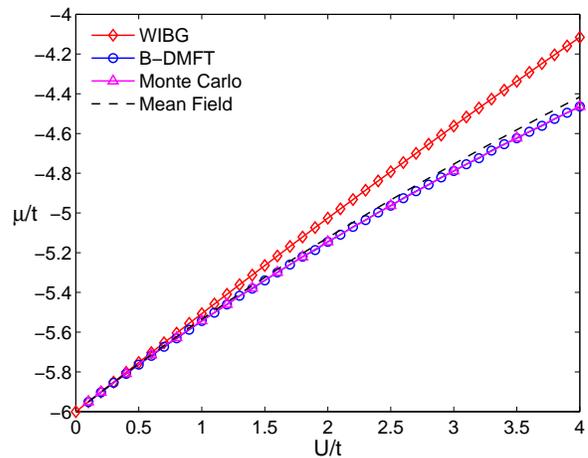}
\caption{(Color online)  Shown is the chemical potential as a function of $U/t$ such that the density is approximately $\langle n \rangle =0.5$ for the case of the weakly interacting Bose gas (WIBG), B-DMFT, quantum Monte Carlo worm-type simulations on a lattice of size $L=10^3$, and static mean-field theory. The temperature is $T/t = 0.25$. For low $U$, both static mean-field theory and B-DMFT approach the weakly interacting Bose gas theory well, while in the intermediate regime B-DMFT is doing a superior job for the chemical potential despite the local approximation for the self energy. }
\label{figure_mu_vs_U_WIBG}
\end{center}
\end{figure}

\begin{figure}[t]
\begin{center}
\includegraphics[angle=0, width=1.0\columnwidth]{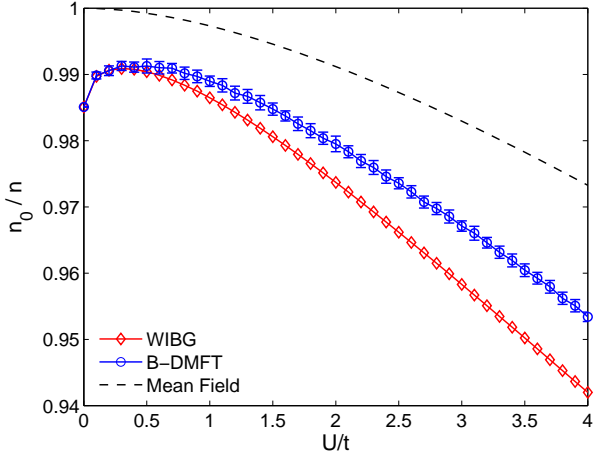}
\caption{(Color online)  Shown is the condensate fraction as a function of $U/t$ such that the density is approximately $ \langle n \rangle =0.5$ for the case of the weakly interacting Bose gas (WIBG), B-DMFT
and static mean-field theory. The temperature is $T/t = 0.25$. For low and intermediate $U$, B-DMFT is in better agreement with the weakly interacting Bose gas than static mean-field theory.}
\label{figure_condfrac_vs_U_WIBG}
\end{center}
\end{figure}

\begin{figure}[ht]
\begin{center}
\includegraphics[angle=0, width=1.0\columnwidth]{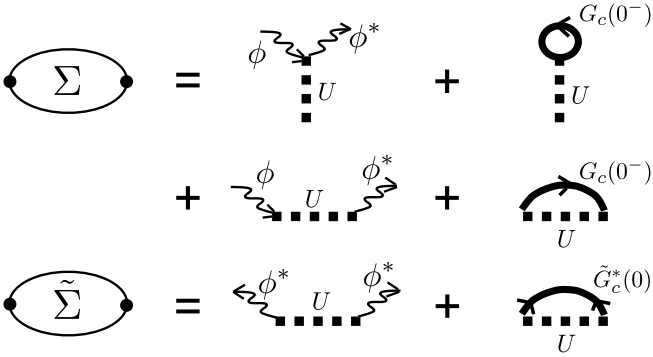}
\caption{First order skeleton diagrams of the selfenergy valid in the weakly interacting Bose gas regime, with results $\Sigma^{(1)} = 2 \langle n_0 \rangle U + 2 G_c(\tau = 0^-)U = 2 \langle n \rangle U$ and $\tilde\Sigma^{(1)} = \langle n_0 \rangle U + \tilde G_c(\tau = 0)U = \langle b b\rangle U$. Note that in single-site B-DMFT all momenta are equal, and our interactions are instantaneous.}
\label{fig:WIBG_diagram}
\end{center}
\end{figure}

In Fig.~\ref{figure_mu_vs_U_WIBG} the chemical potential is calculated such that the density is $\langle n \rangle = 0.5$. The chemical potentials found by B-DMFT are in excellent agreement with the ones found by quantum Monte Carlo worm-type simulations. This shows again that the deviation from the Hugenholtz-Pines relation in B-DMFT is minimal, and goes to zero as $U/t \to 0$. We also compare the condensate density for these theories in Fig.~\ref{figure_condfrac_vs_U_WIBG}, a more challenging quantity than the total density. The agreement with quantum Monte Carlo simulations and the agreement with the WIBG theory for $U/t < 0.5$ is remarkable, and the difference with static mean-field theory is notable.
Technically, calculations in this limit are hampered by the low values of the selfenergies and the proximity of the band edge due to which very precise calculations are needed. For the density  $\langle n \rangle = 0.5$ the sign problem is not the limiting factor (the average sign is approximately  $\langle s \rangle = 0.7$), but the situation deteriorates quickly for higher densities at these temperatures.

To conclude this section, we briefly comment on the difference between the anomalous selfenergy found here ($\tilde\Sigma^{(1)} = \langle bb \rangle U$) and the one found in Ref.~\onlinecite{CapogrossoSansone10} ($\tilde\Sigma^{(1)} = \langle n_0 \rangle U$) for a system in continuous space.
The difference between the two is the diagram given by the convolution of the anomalous Green function with the interaction line, at finite momenta (that is the second diagram for $\tilde\Sigma^{(1)}$ in Fig.~\ref{fig:WIBG_diagram}). 
In Ref.~\onlinecite{CapogrossoSansone10} this diagram was not present in first order for the anomalous selfenergy since it was argued that interactions can to leading order be replaced by their $k=0$ values (Eq. (3.44), for which this diagram is zero. The diagram does contribute to the chemical potential however for non-zero momenta, see Fig. 9 on page~22 of Ref.~\onlinecite{CapogrossoSansone10}. In case of B-DMFT however, we cannot make the distinction that the strength of the interaction is different for small or large momenta. Hence, to leading order there is still a contribution of the anomalous Green function to the anomalous self energy; in other words $U\langle b b \rangle$ is the correct value for the selfenergy in the WIBG limit for B-DMFT to leading order.

\begin{figure}[ht]
\begin{center}
\includegraphics[angle=0, width=1.0\columnwidth]{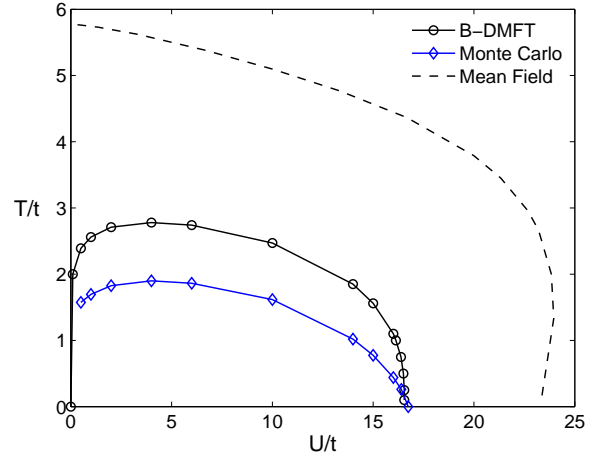}
\includegraphics[angle=0, width=1.0\columnwidth]{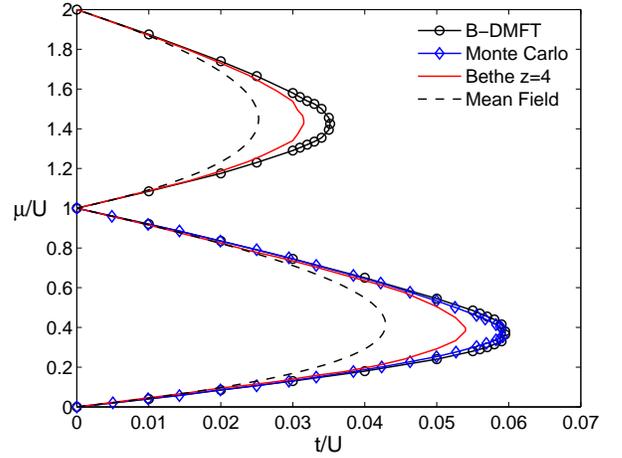}
\caption{(Color online)  Top panel: Phase diagram of the square lattice Bose-Hubbard model in the space of interaction and temperature for $n=1$. Bottom panel: ground-state phase diagram in the space of $t/U$ and $\mu/U$, showing the first two Mott lobes surrounded by superfluid. The B-DMFT phase boundary was computed at $\beta t=8$.  The dashed line shows the static mean-field result, the red curve the exact solution for a Bethe lattice with coordination number $z=4$ (Ref.~\onlinecite{Semerjian09}) and the blue curve with open diamonds the QMC result from lattice simulations (Ref.~\onlinecite{CapogrossoSansone08}). The black line with open circles corresponds to the B-DMFT solution. Error bars are much smaller than the symbol size.}
\label{ground_state_2d}
\end{center}
\end{figure}

\subsection{2d square lattice} 
\label{2d_sec}

\begin{figure}[ht]
\begin{center}
\includegraphics[angle=0, width=1.0\columnwidth]{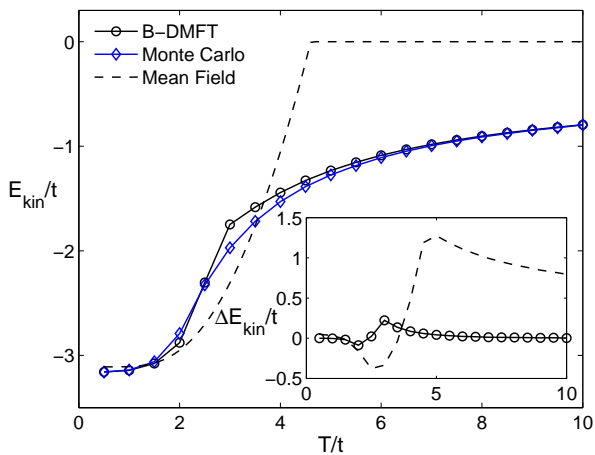}
\caption{(Color online)  Kinetic energy of the Bose-Hubbard model on the 2d square lattice as a function of temperature for $\mu/t=3.52$ and $U/t=10$ ($n\approx1$). Results obtained from B-DMFT (black circles) are compared to lattice QMC (blue diamonds) and to static mean-field theory (black dashed line). Inset: Energy difference from the the QMC data for the same parameters. The QMC results are obtained on a lattice with $100^2$ sites. Error bars are smaller than the symbol size.}
\label{figure_ekin_2d}
\end{center}
\end{figure}

In this section we present results for the 2d square lattice. As in the previous section data from B-DMFT are compared to lattice QMC,\cite{CapogrossoSansone08} static mean-field theory and an exact solution of this model on a Bethe lattice with connectivity $z=4$.\cite{Semerjian09} In Fig.~\ref{ground_state_2d} we show the finite temperature phase diagram (top panel) and the ground state phase diagram (bottom panel) for the first and second lobe on the 2d square lattice. The ground state phase diagram was calculated at $\beta t=8$ which is sufficiently low as can be seen from the top panel of Fig.~\ref{ground_state_2d}. As one might expect, the agreement of B-DMFT with exact Monte Carlo lattice simulations is not as good in 2d as in 3d. We note here that in Fig.~\ref{ground_state_2d} we compare the condensate transition obtained by B-DMFT to the SF transition obtained by Monte Carlo lattice simulations. Since B-DMFT is just a dynamical extension to static mean-field theory we obtain (as in the static case) a finite condensate $\phi$ which is not present in a 2d system except at $T=0$ according to the Hohenberg-Mermin-Wagner theorem. Similar as in the discussion of Ref.~\onlinecite{CapogrossoSansone10} on the weakly-interacting Bose gas theory, Beliaev's diagrammatic technique is useful in computing thermodynamic observables since they converge on short-range distances. The long-range fluctuations of the phase that ultimately destroy the condensate can be described at long wavelengths with a separate hydrodynamical action. The diagrammatic technique simplifies hence enormously by working with explicit symmetry breaking and a genuine condensate in the calculation of thermodynamic observables.~\cite{CapogrossoSansone10} These ideas are illustrated for the kinetic energy in Fig.~\ref{figure_ekin_2d}, where at high and low temperatures the agreement with quantum Monte Carlo worm-type simulations is excellent, in contrast to static-mean field theory. Only in the vicinity of the transition point relatively small deviations are found. As in 3d, B-DMFT predicts a kink in the kinetic energy curve at the transition point, and is hence not well equipped to describe Kosterlitz-Thouless phase transitions.
Still, it gives a substantially improved finite temperature phase diagram compared to static mean-field theory (top panel of Fig.~\ref{ground_state_2d}) and a ground state phase diagram (bottom panel) in good agreement with the QMC result.

\subsection{Bethe lattice}\label{bethe_sec}
\subsubsection{Finite connectivity}

\begin{figure}[t]
\begin{center}
\includegraphics[angle=0, width=1.0\columnwidth]{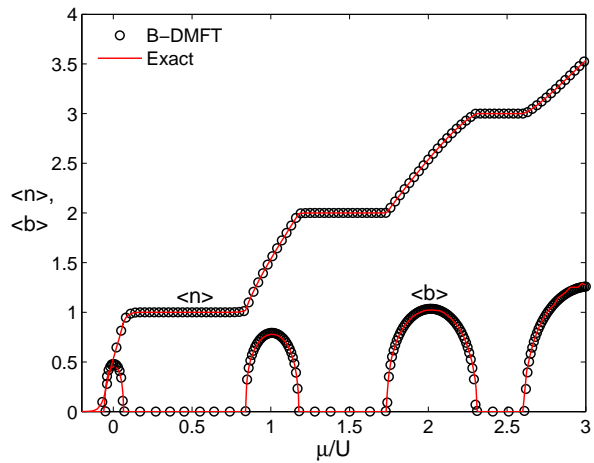}
\caption{(Color online)  Density $n$ and condensate order parameter $\phi=\langle b \rangle$ as a function of $\mu/U$ for $t/U=0.02$ and $\beta t=1$ on the Bethe lattice with connectivity $z=4$. The open black circles are the result obtained with B-DMFT and the solid red line the numerically exact solution (Ref.~\onlinecite{Semerjian09}).}
\label{comparison_bethe_B-DMFT}
\end{center}
\end{figure}

\begin{figure}[t]
\begin{center}
\includegraphics[angle=0, width=1.0\columnwidth]{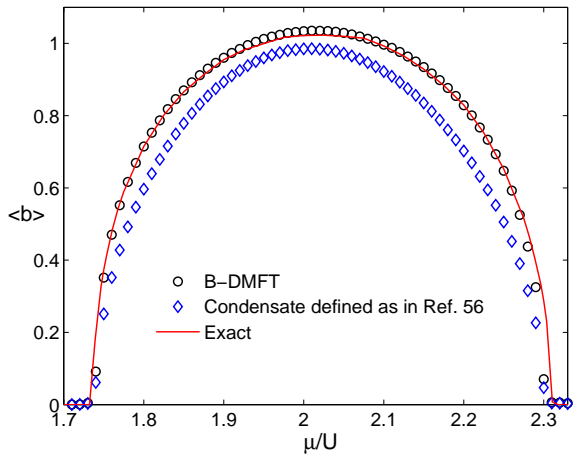}
\caption{(Color online)  Condensate order parameter $\phi=\langle b \rangle$ as a function of $\mu/U$ for $t/U=0.02$ and $\beta t=1$ on the Bethe lattice with connectivity $z=4$ (blow-up of $\phi$ in Fig.~{\ref{comparison_bethe_B-DMFT}} around $\mu/U=2$). The open black circles are the result obtained with B-DMFT, the solid red line the numerically exact solution (Ref.~\onlinecite{Semerjian09}) and the blue diamonds are results for the condensate $\phi$ defined as in Ref.~\onlinecite{Byczuk10}.}
\label{comparison_bethe_B-DMFT2}
\end{center}
\end{figure}

In this section we compare the solution of the B-DMFT equation on the Bethe lattice with the exact solution by  Semerjian and co-workers.\cite{Semerjian09} Since both methods are formulated in the thermodynamic limit 
this allows us to directly compare results and to check the accuracy of the B-DMFT formalism. For the Bethe lattice with connectivity $z$ the density of states is given by
\begin{equation}
D(\epsilon) = \frac{1}{V} \delta( \epsilon + zt) + D_c(\epsilon),
\end{equation}
with $V$ the volume of the system. The continuum part $D_c$ reads
\begin{equation}
D_c(\epsilon) = \frac{\sqrt{4(z-1)t^2 - \epsilon^2}}{2\pi(zt^2-\epsilon^2/z)}, \hspace{1.0cm} |\epsilon| \le 2t\sqrt{z-1},
\label{dos_bethe}
\end{equation}
where $t$ is the hopping between the sites. The isolated state at $\epsilon=-zt$ matters for condensation and prevents the existence of Goldstone modes in the symmetry broken phase.\cite{Laumann09} In Fig.~{\ref{comparison_bethe_B-DMFT}} we show a comparison of the density $n$ and the condensate order parameter $\phi$ between B-DMFT and the exact solution for a Bethe lattice with connectivity $z=4$. As one can see in Fig.~{\ref{comparison_bethe_B-DMFT2}}, the agreement is excellent except for the condensate $\phi$ very close to the SF to Mott transition. This comes from the fact that B-DMFT does not reproduce the mean-field exponents like the exact solution~\cite{Semerjian09} but has a smaller exponent in this case, i.e. the transition is much sharper. This excellent agreement of the B-DMFT condensate $\phi$ with the exact solution also shows that the definition of the condensate used both in this work and in our previous Letter, Ref.~\onlinecite{Anders10}, is correct, unlike the condensate defined in the Comment~\cite{Byczuk10} on our previous work,\cite{Anders10} which corresponds to the blue diamonds in Fig.~\ref{comparison_bethe_B-DMFT2}.

\subsubsection{Limit of infinite connectivity}

For systems without symmetry breaking, the isolated state in the non-interacting density of states is negligible. It is customary in (fermionic) DMFT to rescale the hopping in the limit $z \to \infty$ as  $t = \tilde{t}/\sqrt{z}$ in order for the density of states to remain finite.
Equation~(\ref{dos_bethe}) reduces then to the semi-circular density of states given by

\begin{equation}
D(\epsilon) = \frac{1}{2 \pi \tilde t^2} \sqrt{4\tilde t^2 - \epsilon^2},  \hspace{1.0cm} |\epsilon| \le 2\tilde{t}.
\label{dos_semi}
\end{equation}
For this density of states the selfconsistency equations simplify since the Hilbert-transformation can be performed analytically. This allows us to formulate the whole selfconsistency loop in $\tau$-space, i.e. we do not have to go to Matsubara frequencies. We obtain a simple relation between the hybridization and the Green's function given by

\begin{equation}
\label{eq:Hilbert_for_Bethe}
\mathbf{\Delta}(\tau) = - \tilde t^2 \mathbf{G}_{c}(\tau).
\end{equation}

For bosons, however, the isolated state in the non-interacting density of states is not negligible. As soon as $z > 2$ the condensation temperature is non-zero, and one has then to keep track of the uniform mode.\cite{Laumann09} Integer scaling $t = \tilde{t}/z$ is then the only possibility, and the static mean-field theory~\cite{Fisher89} becomes exact for $z \to \infty$ (for all temperatures). This integer scaling is also compatible with Eq.(\ref{kappa}) such that the hybridization terms become identically zero for $z \to \infty$, again in line with Refs.~\onlinecite{Fisher89, Laumann09} but in contrast with the different scalings postulated in Ref.~\onlinecite{Byczuk08}.



\section{Conclusions}\label{conclusion_sec}

In summary, we have derived the B-DMFT formalism in a number of different ways: (i) by approximating the kinetic energy functional, (ii) by considering a cavity method or $1/z$ expansion, (iii) by using an effective medium approach.
We highlighted similarities and differences with previous approaches.~\cite{Byczuk08, Hu09, Hubener09, Snoek10a, Semerjian09} We discussed the Monte Carlo impurity solver in great detail including detailed balance of the updates, technical difficulties one may encounter, and the measurement procedure. We have shown results for the 3d Bose-Hubbard model, where the zero temperature phase diagram and the finite temperature phase diagram are in excellent agreement with quantum Monte Carlo worm algorithm results. We compared with the theory of the weakly interacting Bose gas in the regime of weak repulsion. The kinetic energy is reproduced at the $1-2\%$ level over the full range of temperatures, except in the very close vicinity to the superfluid-to-normal transition where deviations are larger. We studied the critical exponents of this transition on the superfluid side, and found non-universal values depending on the parameters of the Bose-Hubbard model that are different from the mean-field values and from the universal ones found in the 3d XY universality class. The reason is that the contribution of the hybridization terms can be understood as a one-loop correction to the condensate. In two dimensions, the ground state phase diagram is still in good agreement with quantum Monte Carlo worm simulations, but at finite temperature the superfluid-to-normal transition is less accurately reproduced, because of the Kosterlitz-Thouless nature of the phase transition: in B-DMFT there is still a genuine condensate at finite temperature in 2d, and we miss the long-range fluctuations of the phase. However, quantities that are local such as the energy are still remarkably accurately reproduced over the entire parameter regime, just as in 3d. We also compared B-DMFT with the exact solution of Ref.~\onlinecite{Semerjian09} on the Bethe lattice, and found again good agreement except in the very close vicinity of the phase transition points. 
The agreement over the whole parameter regime (except in the  vicinity of phase transitions) was argued to be the result  of fast diagrammatic convergence of thermodynamic quantities with system size, similarly as for the weakly interacting Bose gas system.~\cite{CapogrossoSansone10} Our approach constitutes thus a natural extension of the Bogoliubov mean-field theory of the weakly interacting system (which is unable to find a Mott insulator) to a dynamical mean-field theory successful over the entire parameter regime of the Bose-Hubbard model. It shows that, when technical difficulties such as the asymptotic behavior of the series are properly taken care of or sidestepped, elusive diagrammatic expansions for bosonic systems are in fact promising, in line with the spirit of (bold) diagrammatic Monte Carlo~\cite{Prokofev_BMC} studies, whether or not in combination with DMFT.~\cite{Pollet10_inc} 

The virtue of developing a DMFT formalism for the Bose-Hubbard model lies in the possible extension of the formalism to models where known numerical methods fail. Prime examples are Bose-Fermi mixtures in two or three dimensions,\cite{Byczuk09, Gunter06, Ospelkaus06, Best09} where previous calculations treated the spinless fermions within DMFT and the bosons at the mean-field level.\cite{Snoek10b} The formalism can also be generalized to mixtures with spinful fermions. Cluster expansions are  possible along the lines of Ref.~\onlinecite{Werner06Kondo}, but it remains to be seen whether the sign problem remains tolerable in the condensed phase. Developing the formalism for real-time applications can be considered along the same lines as for fermionic DMFT,\cite{Eckstein10} but here, too, it remains to be seen how the sign (as well as a complex condensate) behaves.


\acknowledgments
The calculations have been performed on the Brutus cluster at ETH Zurich by using the ALPS libraries \cite{ALPS}. We acknowledge very helpful discussions with F. Assaad and his group, K. Byczuk, A. Georges, G. Kotliar, E. Kozik, C. May, N. Prokof'ev, B. Svistunov, and D. Vollhardt, and thank W.-J. Hu, N.-H. Tong, A. Hubener, M. Snoek, and W. Hofstetter for sharing their B-DMFT results  with us. We are grateful to G. Semerjian and F. Zamponi for sharing their notes on the cavity method with us. 
This project was supported by the Swiss National Science Foundation under Grants PZ00P2-121892, PP002-118866, NSF under Grant No. DMR-1006282, and by a grant from the Army Research Office with funding from the DARPA OLE program. We also acknowledge hospitality of KITP Santa Barbara and the Aspen Center for Physics.


\begin{appendix}

\section{Quantum Cavity Method ($1/z$ expansion)}
\label{sec:cavity}

In a recent paper,~\cite{Semerjian09} Semerjian {\it et al.} derived an exact solution for the Bose-Hubbard model on the Bethe lattice (lattices with tree structure). Their final result is a selfconsistency equation that can easily be solved numerically. The paper mentions that B-DMFT can be considered as the first order approximation to this equation. This is indeed the case, and leads to an equivalent derivation of the B-DMFT equations. The expressions in Ref.~\onlinecite{Semerjian09} are however for the cavity condensate and Green's function, not for the true (observable) condensate and Green's function. In this paragraph we write down their equations for  the observable quantities and compare with our B-DMFT action.\cite{Zamponi}

Computations on a tree can benefit from the recursive structure of the tree. Following Ref.~\onlinecite{Semerjian09}, the quantity $Z_{i\to j} (\{ n \}) $ for two adjacent sites $i$ and $j$ is defined as the partial partition function for the subtree rooted at $i$ and excluding the branch to $j$ for given quantum numbers $\{ n \}$. We note that a Suzuki-Trotter decomposition for the quantum action preserves the structure of the lattice for every time slice. Normalized quantities which can be interpreted as probabilities can then be defined as $\eta_{i \to j} ( \{ n \} )  = Z_{i \to j} ( \{ n \} )  / \sum_ {( \{ n' \} ) }   Z_{i \to j} ( \{ n' \} )$. We refer to Ref.~\onlinecite{Semerjian09} for more details.

When taking an infinite number of Suzuki-Trotter slices, the cavity field on a Bethe lattice of connectivity $z$ can selfconsistently be expressed as~\cite{Semerjian09}
\begin{widetext}
\begin{equation}
\eta_{\rm cav} (b^*, b) = \frac{1}{Z_{\rm cav}} w(b^*, b) \int  \prod_{i=1}^{z-1} \mathcal{D}[ b_i^*, b_i] \eta_{\rm cav}(b_i^*, b_i) \exp \left[ t \int_0^{\beta} d\tau  \left( b^*(\tau) \sum_{i=1}^{z-1} b_i(\tau) + b(\tau) \sum_{i=1}^{z-1} b^*_i(\tau)  \right)  \right] ,
\end{equation}
with the on-site weight of the path given by
\begin{equation}
w(b^*, b) = \exp  \left[ - \int_0^{\beta} d\tau \left( b^*(\tau) (\partial_{\tau} - \mu ) b(\tau) + \frac{U}{2} n(\tau) (n(\tau) - 1) \right) \right].
\end{equation}
\end{widetext}
In Nambu notation this reads
\begin{eqnarray}
\eta_{\rm cav} (\textbf{b})  & = & \frac{1}{Z_{\rm cav}} w(  \textbf{b} ) e^{(z-1) \Gamma(t \textbf{b} ) }, \nonumber \\
\Gamma( \mathbf{\Phi} ) & = & \ln \left[ \int {\mathcal D} \textbf b \eta_{\rm cav}(\textbf b) \exp \int_0^{\beta} d\tau \textbf{b}^{\dagger} (\tau) \mathbf{\Phi} (\tau) \right]. \hspace{5mm}
\label{eq:Zamponi_eta}
\end{eqnarray}
This selfconsistency equation can only be solved exactly for $U=0$. For the interacting case, $U > 0$, a solution can be looked for in the large connectivity limit, where in the leading order $1/z$ B-DMFT is recovered. This follows from an expansion of the generating functional of connected correlation functions $\Gamma$, 
\begin{align}
&\Gamma({\mathbf \Phi} ) =  \int_0^{\beta} d\tau \langle \textbf{b}^{\dagger} \rangle_{\rm cav} \mathbf{\Phi}(\tau) \nonumber \\
&\hspace{8mm}+  \frac{1}{2} \int_0^{\beta} d\tau d\tau' \mathbf{\Phi}^{\dagger} (\tau) {\mathbf G}^c_{\rm cav} (\tau - \tau')  \mathbf{\Phi}(\tau') + \ldots .
\label{eq:zamponi_gamma}
\end{align}
The connected part of the two-point correlator satisfies
\begin{equation}
{\mathbf G}^c_{\rm cav} (\tau - \tau') = \langle \mathbf{b}^{\dagger} (\tau)  \mathbf{b}(\tau') \rangle_{\rm cav} - \langle \mathbf{b}^{\dagger} (\tau) \rangle_{\rm cav} \langle  \mathbf{b}(\tau') \rangle_{\rm cav}.
\end{equation}
%
We now plug Eq.~(\ref{eq:zamponi_gamma}) into Eq.~(\ref{eq:Zamponi_eta}) and obtain
\begin{align}
&\eta_{\rm cav}( \textbf{b} )  =  \frac{1}{Z_{\rm cav}} \exp \left[ - S_{\rm cav} \right] , \nonumber \\
&S_{\rm cav}  =   \frac{1}{2} \int_0^{\beta} d\tau d\tau' \textbf{b}^{\dagger}(\tau) \textbf{G}_{0, {\rm cav}}^{-1}(\tau - \tau') \textbf{b}(\tau') \nonumber \\
& \hspace{10mm}+ \int_0^{\beta} d\tau \left[ \frac{U}{2} n(\tau) ( n(\tau) - 1) - t (z-1) \mathbf{\Phi}^{\dagger}_{\rm cav} \textbf{b} (\tau) \right] , \nonumber \\
&\textbf{G}_{0, {\rm cav}}^{-1} (\tau - \tau')  =   (\partial_{\tau} \sigma_3 - \mu \mathbf{I}) \delta(\tau - \tau')\nonumber\\
& \hspace{26mm} - t^2 (z-1) {\mathbf G}^c_{\rm cav} (\tau - \tau').
\label{eq:Zamponi_cavity_action}
\end{align}
In the limit of large connectivity, one has to scale the hopping as $t \sim 1/z$, hence $zt \sim 1$ and $t^2z ~\sim 1/z$. \\

We now proceed to the equations for the true fields, which cannot be found in Ref.~\onlinecite{Semerjian09}.\cite{Zamponi} To this end, the true marginal $\eta( \textbf{b} )  =  \frac{1}{Z} w( \textbf{b}) e^{z \Gamma(t \textbf{b}) }$ is needed. It is still expressed in terms of the cavity field (and not the true field) and is obtained by replacing $z-1$ by $z$ in Eq.~(\ref{eq:Zamponi_cavity_action}). We try to find now  suitable expressions for the true connected Green's function and condensate, valid for the impurity problem `imp' instead of for the cavity problem `cav'. The action for the impurity problem is valid up to order $1/z$. Since the prefactor of the cavity Green's function is already of ${\mathcal{O}}(1/z)$, we can identify the impurity Green function with the cavity Green function without loss of accuracy,
\begin{eqnarray}
{\mathbf G}^c_{\rm imp} (\tau - \tau') &  = & {\mathbf G}^c_{\rm cav} (\tau - \tau')  \nonumber \\
{} & = &  \langle \mathbf{b}^{\dagger} (\tau)  \mathbf{b}(\tau') \rangle_{\rm imp} - \langle \mathbf{b}^{\dagger} (\tau) \rangle_{\rm imp} \langle  \mathbf{b}(\tau') \rangle_{\rm imp},\nonumber\\
\end{eqnarray}
where the average can now be taken with respect to the impurity action.

The prefactor of the condensate in Eq.~(\ref{eq:Zamponi_cavity_action}) is of order unity, hence we need the impurity condensate to ${\mathcal{O}} (1/z)$. We have
\begin{equation}
S_{\rm imp} = S_{\rm cav} - t \int_0^{\beta} d\tau \mathbf{\Phi}^{\dagger} \textbf{b}(\tau).
\label{eq:app_action_imp}
\end{equation}

The condensate is now found for the total impurity action $S_{\rm imp}$,
\begin{eqnarray}
\mathbf{\Phi}_{\rm imp} & = & \langle \textbf{b} \rangle_{\rm imp} \nonumber \\
{} & \sim & \mathbf{\Phi}_{\rm cav} + t \int_0^{\beta} d\tau {\mathbf G}^c_{\rm imp} (-\tau )  \mathbf{\Phi}_{\rm cav},
\end{eqnarray}
which can be inverted (accurate up to this order),
\begin{eqnarray}
 \mathbf{\Phi}_{\rm cav}  \sim \left( \mathbf{I} - t {\mathbf G}^c_{\rm imp}(i \omega_n = 0 ) \right) \mathbf{\Phi}_{\rm imp}.
\end{eqnarray}
By plugging this equation into Eq.~\ref{eq:app_action_imp} a closed set of equations for the true condensate is found, which is tantamount to B-DMFT. The iteration scheme proceeds in the usual way.
Note that the last step does not rely on the scaling in Eq.~(\ref{eq:Hilbert_for_Bethe}) and Eq.(\ref{dos_semi}), but does use the recurrence relation of a tree.
We note that the derivation presented in Ref.~\onlinecite{Snoek10a} is conceptually equivalent to the one presented in Ref.~\onlinecite{Semerjian09}, on which this appendix is based.


\section{Effective medium approach}\label{medium_sec}
In this section we will expand the full Bose-Hubbard action around a single site, identify the low-energy modes, and re-exponentiate them.
The derivation in this section is similar to the derivation found in Ref.~\onlinecite{Byczuk08}, but differs in the treatment of the condensate. We provide a microscopic RG-like picture for the condensate field and hybridization fields in the effective action. The total action of the full Bose-Hubbard model is split into three parts,
with the first one being the local part defined in Eq.~(\ref{eq:action_internal}),
\begin{equation}
S_{\rm int} = \int_0^{\beta} d\tau b_{\rm int} ^*(\tau) ( \partial_{\tau} - \mu) b_{\rm int} (\tau) + \frac{U}{2}   n_{\rm int} (n_{\rm int}-1).
\end{equation}
The second part $\Delta S$ describes the coupling with the rest of the lattice,
\begin{equation}
\Delta S = \int_0^{\beta} \Delta S(\tau) = \int_0^{\beta}  \sum_{\langle {\rm int}, {\rm ext} \rangle} -t \left( b^{\dagger}_{\rm int} b_{\rm ext} + b^{\dagger}_{\rm ext} b_{\rm int} \right)
\label{eq:action_coupling}
\end{equation}
and the third part is the remaining, external part `ext' with action
\begin{equation}
S_{\rm ext} =  \int_0^{\beta} d\tau b_{\rm ext} ^*(\tau) ( \partial_{\tau} - \mu) b_{\rm ext} + \frac{U}{2}   n_{\rm ext} (n_{\rm ext}-1).
\end{equation}
We had that $ S = S_{\rm int} + S_{\rm ext} + \Delta S$, but would like to derive an action $S \approx S_{\rm imp} + S_{\rm ext}$ by expanding the exponential function with respect to the action $\Delta S$, and perform the functional integral over the external degrees of freedom, and incorporate this into a new effective internal action $S_{\rm imp}$ and a decoupled, external action $S_{\rm ext}$.

We assume that `ext' is a thermodynamic system that can spontaneously break the symmetry, if needed, and develop a condensate. If so, we decompose
\begin{equation}
b_{\rm ext} = \phi_{\rm ext} + \delta b_{\rm ext},
\end{equation}
where $ \langle b_{\rm ext} \rangle = \phi_{\rm ext}$ represents the condensate wave function. It is a classical field ($c$-number, taken to be real) that breaks the symmetry, and its presence forces us to consider anomalous averages such as $ \langle b_{\rm ext} b_{\rm ext} \rangle$, which are nonzero.
The commutation relations are now $[ \delta b_{\rm ext}^{(j)} , \delta b^{\dagger (k) }_{\rm ext}] = \delta_{j,k}$ and zero otherwise.
The impurity part `int' is by itself small and cannot spontaneously break the symmetry and develop a non-zero expectation value for its local operators. \\
In our perturbative expansion in $\Delta S$ an atom from the `int' part will hop to the `ext' part and acquire an expectation value through the full correlators in `ext'.
A non-zero expectation value corresponds to the condensate mode, and is not possible for the bosons outside the condensate. We do not attempt here to build in all low-energy modes via `ext', only the classical field contributions from the condensate will suffice.
Let us anticipate in advance on such a possibility and allow
\begin{equation}
b_{\rm int} = \phi_{\rm int} + \delta b_{\rm int}, \quad \langle b_{\rm int} \rangle = \phi_{\rm int}.
\label{eq:expect_value_int}
\end{equation}
We also have the commutation relations  $[ \delta b_{\rm int}^{(j)} , \delta b^{\dagger (k) }_{\rm int}] = \delta_{j,k}$ and $[ \delta b_{\rm int}^{(j)} , \delta b^{\dagger (k) }_{\rm ext}] = 0$.
In second order, we then also have to allow for $\langle b_{\rm int} b_{\rm int} \rangle$ to become large.
The validity of Eq.~(\ref{eq:expect_value_int}) can then be checked {\it a posteriori}.

The coupling between the internal and external degrees of freedom, expressed in  Eq.~(\ref{eq:action_coupling}), can then be written as
\begin{eqnarray}
\Delta S & = &  \int_0^{\beta}  \sum_{\langle {\rm int}, {\rm ext} \rangle} -t \left( b^{\dagger}_{\rm int} b_{\rm ext} + b^{\dagger}_{\rm ext} b_{\rm int} \right) \nonumber \\
{} & = &  -t \int_0^{\beta} d\tau z  \phi_{\rm int} ( \delta b_{\rm ext} + \delta b^{\dagger}_{\rm ext} )  + z \phi_{\rm ext} ( \delta b_{\rm int} + \delta b^{\dagger}_{\rm int} )  \nonumber \\
{} & {} & +   \sum_{\langle {\rm int}, {\rm ext} \rangle} \delta b^{\dagger}_{\rm int} \delta b_{\rm ext} + \delta b^{\dagger}_{\rm ext} \delta b_{\rm int},
\end{eqnarray}
where we have omitted terms in $\phi_{\rm int} \phi_{\rm ext}$ since the condensate is treated as a $c$-number and such terms involve an arbitrary shift of the condensate energy. There are four terms in $\Delta S$, two of which have the condensate directly in them, and two that don't. Since we are interested in adding the physics of the WIBG to the action,
we have to treat the condensate as a large contribution compared to the small contributions coming from the depleted atoms, while the third term will be contributing to one (and higher) loop corrections. 
We can now immediately write down the contribution of the parts with the condensate lines to the action,
\begin{equation}
S_1 = zt \phi_{\rm ext} \int \delta b(\tau) + \delta b^{\dagger}(\tau) = zt \phi_{\rm ext} \int b(\tau) + b^{\dagger}(\tau)
\end{equation}
and we keep only the following terms in $\Delta S$ (the other one factorizes), which we treat as a small term,
\begin{equation}
\Delta S =  -t \int_0^{\beta} \sum_{\langle {\rm int}, {\rm ext} \rangle} \delta b^{\dagger}_{\rm int} \delta b_{\rm ext} + \delta b^{\dagger}_{\rm ext} \delta b_{\rm int}.
\end{equation}
The advantage of our approach is that we have first added the condensate to the impurity action, before looking at depletion (fluctuation) terms. In this picture, the stationarity and time-independence of the condensate is immediately built in.
We are now in a position to expand in $\Delta S$.
This results in an infinite series,
\begin{eqnarray}
Z & \sim  & Z^{\rm ext} \int D [b_{\rm int}^*, b_{\rm int} ] e^{-S_{\rm int}  [b_{\rm int}^*, b_{\rm int} ] - S_1  [b_{\rm int}^*, b_{\rm int} ]}  \zeta \nonumber \\
\zeta & = & 1 - \int_0^{\rm \beta} d\tau \langle \Delta S(\tau) \rangle_{\rm ext} + \nonumber \\
{} & {} & + \frac{1}{2}  \int_0^{\rm \beta} d\tau_1  \int_0^{\rm \beta} d\tau_2 \langle \Delta S(\tau_1) \Delta S(\tau_2) \rangle_{\rm ext} + \ldots .
\end{eqnarray}
We start with the evaluation of the first order term,
\begin{equation}
\int_0^{\beta} \langle \Delta S \rangle _{\rm ext}  = -t \int_0^{\beta} d\tau \sum_{\langle {\rm int, ext} \rangle} \langle \delta b_{\rm int}^{\dagger} \delta b_{\rm ext} \rangle_{\rm ext} + {\rm h.c.} ,
\end{equation}
which is zero, as could have been expected from symmetry. \\

The second order term is the lowest order term that survives,
\begin{eqnarray}
\frac{1}{2}  \int_0^{\rm \beta} d\tau_1  \int_0^{\rm \beta} d\tau_2 \langle \Delta S(\tau_1) \Delta S(\tau_2) \rangle_{\rm ext} = \nonumber \\
 \frac{t^2}{2}  \int_0^{\rm \beta} d\tau_1  \int_0^{\rm \beta} d\tau_2 \sum_{j,k \in {\rm ext}} [ S^1 + S^2 +S^3 + S^4 ].
\end{eqnarray}
with $S^j, j = 1, \ldots, 4$,
\begin{eqnarray}
S^1 & = & \langle \delta b_{\rm int} (\tau_1) \delta b_{\rm int} (\tau_2) \delta b_{\rm ext }^{\dagger (j)} (\tau_1) \delta b_{\rm ext}^{\dagger (k) } (\tau_2) \rangle_{\rm ext}, \nonumber \\
S^2 & = & \langle \delta b_{\rm int} (\tau_1) \delta b_{\rm int}^{\dagger} (\tau_2) \delta b_{\rm ext }^{\dagger (j)} (\tau_1) \delta b_{\rm ext}^{(k)} (\tau_2) \rangle_{\rm ext}, \nonumber \\
S^3 & = & \langle \delta b_{\rm int}^{\dagger} (\tau_1) \delta b_{\rm int} (\tau_2) \delta b_{\rm ext }^{(j)} (\tau_1) \delta b_{\rm ext}^{\dagger (k) } (\tau_2) \rangle_{\rm ext}, \nonumber \\
S^4 & = & \langle  \delta b_{\rm int}^{\dagger} (\tau_1) \delta b_{\rm int}^{\dagger} (\tau_2) \delta b_{\rm ext }^{(j)} (\tau_1) \delta b_{\rm ext}^{(k) } (\tau_2) \rangle_{\rm ext}. 
\end{eqnarray}
The notation $(j)$ and $(k)$ denotes different sites that are coupled to the impurity through the hopping term in the Bose-Hubbard Hamiltonian.
The anomalous terms survive in the presence of a condensate, which tells us that there are four terms in second order.
Let us analyze $S^1$ in more detail,
\begin{equation}
S^1 = \delta b_{\rm int} (\tau_1) \delta b_{\rm int} (\tau_2) \langle  \delta b_{\rm ext }^{\dagger (j)} (\tau_1) \delta b_{\rm ext}^{\dagger (k)} (\tau_2)  \rangle_{\rm ext}. 
\end{equation}
This describes the anomalous process of a depleted boson hopping from the impurity site to the environment, propagating there according to the full (but unknown to us) anomalous two-particle Green's function, and finally being annihilated on the impurity site.
We will now perform a cumulant re-exponentiation of this term in the presence of a condensate, for which only the connected diagrams can be kept,
\begin{equation}
e^{-S_2^1} =  e^{ - \int_0^{\beta} d\tau_1 \int_0^{\beta} d\tau_2 \delta b_{\rm int} (\tau_1) K(\tau_1 - \tau_2) \delta b_{\rm int}  (\tau_2)},
\end{equation}
where $K$ will be a function which we treat as unknown and which originates from the connected diagrams only. The DMFT selfconsistency iteration scheme will determine $K$ selfconsistently.
In a numerical algorithm it is practical to work with full operators $b$ instead of $\delta b$, so we will write $ b_{\rm int} (\tau) - \phi_{\rm int} $ instead of $\delta b_{\rm int}$.
The other terms $S^j, j = 2,3,4$ are treated in the same way as $S^1$, but we will not write this out explicitly.

%

After convergence, we wish that our site is in equilibrium with the surroundings. We thus impose a first selfconsistency relation
\begin{equation}
\phi = \phi_{\rm ext} = \phi_{\rm int} = \langle b \rangle_{\rm int}.
\end{equation}
The meaning of this relation is that in every iteration we have to compute $\langle b \rangle$ and equate the condensate with this value. It can always be taken real and time-independent.
We collect now all terms needed for our effective impurity action,
\begin{widetext}
\begin{eqnarray}
S_\text{imp} & = &  \int_0^{\beta} d\tau b_{\rm int} ^*(\tau) ( \partial_{\tau} - \mu) b_{\rm int}(\tau) + \frac{U}{2}   n_{\rm int} (n_{\rm int}-1)  - zt \mathbf{\Phi}^{\dagger}_{\rm ext} \int_0^{\beta} d\tau  \textbf{b}_{\rm int}(\tau)  \nonumber \\
{} & {} & + \frac{1}{2} \int_0^{\beta} d\tau  \int_0^{\beta} d\tau' ( \textbf{b}^{\dagger}_{\rm int} (\tau) - \mathbf{\Phi}_{\rm int}(\tau) )   \mathbf{\Delta} (\tau-\tau') ( \textbf{b}_{\rm int} (\tau')  - \mathbf{\Phi}_{\rm int} (\tau') ),
\end{eqnarray}
\end{widetext}
where the elements of the hybridization matrix are (dropping the subscript `c' for connected).
The remaining hybridization term can now be treated similarly as the hybridization term in fermionic DMFT, and has hence a similar selfconsistency relation~\cite{Georges96} (see Sec.~\ref{sec:DMFT_procedure}).


\section{High frequency expansions}
\label{sec:highf}
Within our selfconsistency loop we need to perform Fourier transforms between the imaginary time $\tau$ domain and Matsubara frequencies $\omega_n$. Even though we use a continuous time method we measure the Green's function $\mathbf{G}(\tau)$ on a grid with a fixed number of time slices. We can therefore only obtain accurate data for a finite number of Matsubara frequencies. In order to perform accurate Fourier transforms we supplement the Green's function with the analytically known high-frequency behavior, given by 

\begin{equation}
G(i\omega_n) = \frac{g_1}{i\omega_n} - \frac{g_2}{(i\omega_n)^2} + \frac{g_3}{(i\omega_n)^3} + \Delta G(i\omega_n)
\end{equation}
where $\Delta G(i\omega_n)$ goes to zero as $1/(i\omega_n)^4$ for $i\omega_n\to\infty$. For the static component ($n=0$) we let $\Delta G(i\omega_n)=G(i\omega_n)$. Computing these coefficients analytically improves the accuracy of our Fourier transformation significantly.\cite{Blumer02} By using 
\begin{eqnarray}
\frac{1}{\beta}\sum_{n\ne0} \frac{e^{- i\omega_n \tau}}{i\omega_n} &=& \frac{1}{2\beta}(2\tau-\beta), \nonumber \\
\frac{1}{\beta}\sum_{n\ne0} \frac{e^{- i\omega_n \tau}}{(i\omega_n)^2} &=& \frac{1}{12\beta}(-6\tau^2+6\tau\beta-\beta^2), \nonumber \\
\frac{1}{\beta}\sum_{n\ne0} \frac{e^{- i\omega_n \tau}}{(i\omega_n)^3} &=& \frac{1}{12\beta}(2\tau^3-3\beta\tau^2+\beta^2\tau),
\end{eqnarray}
we can write the inverse Fourier transformation as
\begin{eqnarray}
G(\tau) &=& \frac{1}{\beta}\sum_n G(i\omega_n)e^{-i\omega_n \tau}\nonumber \\
&=& g(\tau) + \frac{1}{\beta}\sum_n \Delta G(i\omega_n)e^{-i\omega_n \tau},
\end{eqnarray}
where
\begin{eqnarray}
g(\tau) &=& \frac{g_1}{2\beta}\,(2\tau-\beta) - \frac{g_2}{12\beta}\,(-6\tau^2+6\tau\beta-\beta^2)\nonumber\\
&+& \frac{g_3}{12\beta}\,(2\tau^3-3\beta\tau^2+\beta^2\tau),
\end{eqnarray}
and the Fourier transformation as
\begin{equation}
G(i\omega_n) =  \frac{g_1}{i\omega_n} - \frac{g_2}{(i\omega_n)^2} + \frac{g_3}{(i\omega_n)^3} + \int_0^{\beta} \Delta G(\tau) e^{i\omega_n \tau},
\end{equation}
with $\Delta G(\tau)=G(\tau)-g(\tau)$.

The coefficients for the Green's function are obtained by evaluating commutators:
\begin{eqnarray}
g_1 &=& \langle[b,b^{\dagger}]\rangle = 1,\nonumber \\
g_2 &=& \langle[[H,b],b^{\dagger}] \rangle = \mu -2U\langle n \rangle, \nonumber \\
g_3 &=& \langle[[H,[H,b]],b^{\dagger}] \rangle \nonumber\\ &=& \langle \epsilon^2\rangle +\mu^2 + 3U^2\langle n^2\rangle -\langle n\rangle (4\mu U+U^2), \nonumber \\
\tilde g_1 &=& \langle[b,b]\rangle = 0, \nonumber \\
\tilde g_2 &=& \langle[[H,b],b] \rangle = U \langle bb \rangle, \nonumber \\
\tilde g_3 &=& \langle[[H,[H,b]],b] \rangle = 0,
\end{eqnarray}
where $g_i$ are the coefficients for the diagonal and $\tilde g_i$ for the off-diagonal Green's function.
All odd coefficients of the anomalous functions are zero, as these functions are purely real since all anomalous functions are symmetric in imaginary time.

Similar coefficients can be obtained for the hybridization function $\Delta(i\omega_n)$. Here we find
\begin{eqnarray}
f_1 &=& \langle \epsilon^2 \rangle, \nonumber \\
f_2 &=& -\langle \epsilon^2\rangle g_2, \nonumber \\
f_3 &=& \langle \epsilon^4\rangle + \langle \epsilon^2\rangle(g_3-2\langle \epsilon^2\rangle),
\end{eqnarray}
for the diagonal component $F(i\omega_n)$ and
\begin{eqnarray}
k_1 &=& 0, \nonumber \\
k_2 &=& -0.5\langle \epsilon^2\rangle \tilde g_2, \nonumber \\
k_3 &=& 0,
\end{eqnarray}
for the off-diagonal component $K(i\omega_n)$.

\end{appendix}


\begin{thebibliography}{99}
\bibitem{Zamponi} The presentation in this section is based on private discussions with the authors of Ref.~\onlinecite{Semerjian09}, as well as on private notes by F. Zamponi, which they allowed us to use in order to clarify this minor issue.
\end{thebibliography}
\end{document}